\documentclass[prb,twocolumn,showpacs,amsmath,amssymb]{revtex4-1}

\usepackage[dvipdfmx]{graphicx}
\usepackage{bm}

\usepackage{color}
\usepackage{braket}
\usepackage{physics}

\newcommand{\thop}{{t_\mathrm{h}}}
\newcommand{\PT}{$PT$}
\newcommand{\calT}{\mathcal{T}}
\newcommand{\calR}{\mathcal{R}}
\newcommand{\kr}{k^\mathrm{r}}
\newcommand{\ki}{k^\mathrm{i}}

\begin{document}

\title{Non-Hermitian Fabry-Perot Resonances in a \PT-symmetric system}

\author{Ken Shobe}
\affiliation{Department of Semiconductor Electronics and Integration Science, 
AdSM, Hiroshima University, 739-8530, Japan}

\author{Keiichi Kuramoto}
\affiliation{Program of Electronic Devices and Systems,
School of Engineering, Hiroshima University, 739-8527, Japan}

\author{Ken-Ichiro Imura}
\affiliation{Graduate School of Advanced Science and Engineering, Hiroshima University, 739-8530, Japan}

\author{Naomichi Hatano}
\affiliation{Institute of Industrial Science, The University of Tokyo, 277-8574, Japan}

\date{October 30, 2020}

\begin{abstract}
In non-Hermitian scattering problems
the behavior of the transmission probability is very different from
its Hermitian counterpart;
it can exceed unity or even be divergent, since the non-Hermiticity can add or remove the probability to and from the scattering system.
In the present paper, we consider the scattering problem of a \PT-symmetric potential and find a counter-intuitive behavior.
In the usual \PT-symmetric non-Hermitian system, we would typically find stationary semi-Hermitian dynamics in a regime of weak non-Hermiticity but observe instability once the non-Hermiticity goes beyond an exceptional point. 
Here, in contrast, the behavior of the transmission probability is strongly non-Hermitian in the regime of weak non-Hermiticity with divergent peaks,
while it is superficially Hermitian 
in the regime of strong non-Hermiticity,
recovering the conventional Fabry-Perot-type peak structure.
We show that the unitarity of the $S$-matrix is generally broken in both of the regimes, but is recovered in the limit of infinitely strong non-Hermiticity.
\end{abstract}

\maketitle

\section{Introduction}

Non-Hermitian quantum mechanics attracts much attention recently; for a recent review, see \textit{e.g.} Ref.~\onlinecite{rev1}.
Historically, it dates back to mid-twentieth century, when nuclear physicists, particularly Feshbach, introduced the idea of optical potential~\cite{opt1,opt2,opt3} for the description of nuclear decay in terms of resonant states with complex eigenvalues in scattering theory.
Feshbach later justified~\cite{fesh1,fesh2}
the complexity of the optical potential by means of projection operators, which we can refer to as a theory of open quantum systems~\cite{oqs1,oqs2,oqs3,Breuer2007} in the present-day terminology.

Interest in non-Hermitian systems was revived in late nineteen-nineties.
A tight-binding model with asymmetric hopping, namely an imaginary vector potential, was introduced in 1996 as an effective model of type-II superconductors, and was connected to the Anderson localization~\cite{hatano1,hatano2}.
This stimulated the theory of non-Hermitian random matrices~\cite{nhrm1,nhrm2,nhrm3}.

A model of oscillator with the parity-time (\PT) symmetric non-harmonic potential was introduced in 1998, originally in order to replace the concept of the Hermiticity as a condition for the reality of the energy eigenvalue~\cite{bender1,bender2}.
This, however, triggered experimental studies on various effectively \PT-symmetric systems,
mainly in the optical ones~\cite{macris,macris_opt,klaiman,guo,ruter,regens,hodaei,feng}, 
and the last couple of years has seen an explosive development of study on non-Hermitian systems of various nature.
The study of \PT-symmetric systems also hinted to generalize arguments on symmetry 
and topology of Hermitian systems~\cite{sato,hughes,gong,KK1} to non-Hermitian ones.
The non-Hermitian topological insulator has been studied both in \PT -symmetric 
~\cite{hughes,schomerus,yuce,poli,weimann,bandres}
and in asymmetric hopping models.
\cite{nhti0,nhti1,nhti2,nhti3,nhti4}

\begin{figure}
\includegraphics[width=70mm]{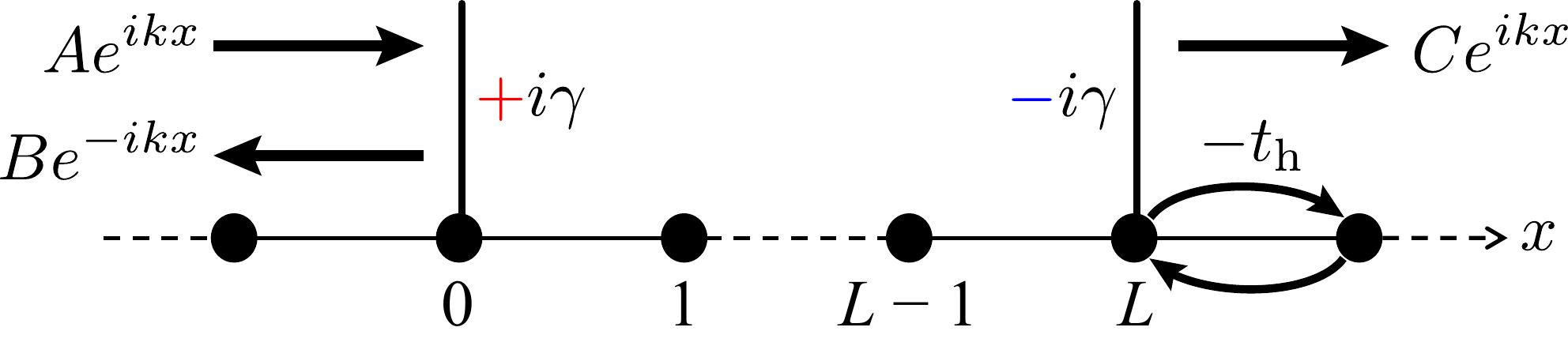}
\\
\caption{Schematic illustration of the scattering problem of a \PT-symmetric model considered in the present paper.
}
\label{schema}
\end{figure}

Yet, fully quantum-mechanical realization of \PT-symmetric systems 
is on its half way~\cite{hide}.
In the present paper, we solve the scattering problem of a \PT-symmetric system shown in Fig.~\ref{schema}, from a perspective of finding a fully quantum-mechanical experimental situation for detecting signatures of the {\PT} symmetry and the non-Hermiticity.
The \PT-symmetric potential may be materialized by attaching environmental systems of source and sink, as is suggested by the studies of optical potential.

A lesson that we can learn from the studies on nuclear physics in the previous century is that the infinite space outside the scatterer in the typical potential scattering problem is in reality terminated by macroscopic neutron injectors and detectors.
Condensed-matter physicists may be more familiar with the same concept in a different context of the Landauer formula~\cite{landauer}.
The infinite leads substitute the macroscopic source and drain to measure the electronic conduction of a microscopic system.
In this sense, any realistic experimental situations are open systems in which macroscopic probes may be represented by the infinite space and the corresponding measurements may be described by solving the scattering problem in an infinite space.
This motivates us to study the scattering problem of a non-Hermitian system from the perspective of measuring the conductance of a \PT-symmetric system.
The non-Hermitian scattering problem for a \PT-symmetric model has been  discussed in different contexts~\cite{PT_scat1,PT_scat2,garmon,mexico,mexico2}.

We here find the following counter-intuitive result:
(i) when the non-Hermitian scattering potential is weak, the non-Hermitian signature is strong in that the transmission probability continuously exceeds unity and occasionally diverges;
(ii) when the non-Hermitian scattering potential goes beyond a threshold
the Hermiticity is seemingly recovered in that the transmission probability shows a Fabry-Perot-type peak structure with all the peaks being less than unity.

In order to contrast the present result with the standard behavior of \PT-symmetric systems, let us briefly review the solution of a prototypical \PT-symmetric model prescribed by the following two-site Hamiltonian:
\begin{align}
H:=
\mqty(
i\gamma  & \thop\\
\thop  &  -i\gamma
),
\label{H_PT2}
\end{align}
where $\thop$ represents the amplitude of hopping between the two sites,
while
$\pm i\gamma$ give a pair of 
two imaginary potentials compatible with {\PT}  symmetry.
The spectrum of Eq.~\eqref{H_PT2} is given by
\begin{align}
E:=\pm \sqrt{\thop^2-\gamma^2};
\end{align}
they are both real in the regime of weak non-Hermiticity $\abs{\gamma}<\abs{\thop}$, which is often called the \PT-unbroken phase, while they are both imaginary in the regime of strong non-Hermiticity $\abs{\gamma}>\abs{\thop}$, which is referred to as the \PT-broken phase.
The Hamiltonian $H$ in Eq.~\eqref{H_PT2} commutes with the {\PT} operator as in 
\begin{align}
(\hat{P}\hat{T})H(\hat{P}\hat{T})^{-1} = H,
\end{align}
or $[H,\hat{P}\hat{T}]=0$, where $P$ is the parity operator
\begin{align}
\hat{P}:=\mqty(
0 & 1 \\
1 & 0
)
\end{align}
and $\hat{T}$ is the time-reversal operator, which in the present case is the complex conjugation. Nonetheless, the eigenstates of $H$ may not be the simultaneous eigenstates of $PT$ because $T$ is an anti-linear operator~\cite{bender05}. 
They are indeed the simultaneous eigenstates in the \PT-unbroken phase, whereas not in the \PT-broken one;
an eigenstate $\ket{\psi_n}$ is parallel to $\hat{P}\hat{T}\ket{\psi}$ in the former, while the two states are the respective eigenvectors of the imaginary eigenvalues $E=\pm i \sqrt{\gamma^2-\thop^2}$ in the latter.

The boundary between the two phases is marked with an exceptional point $\abs{\gamma}=\abs{\thop}$, at which the two eigenvectors of Eq.~\eqref{H_PT2} become parallel to each other and the corresponding eigenvalues coalesce~\cite{Heiss_2012}.
Note that this is distinctive of non-Hermitian systems;
all eigenvectors would be perpendicular to each other in Hermitian systems even when the eigenvalues are degenerate.

This example demonstrates that a non-Hermitian \PT-symmetric system changes its nature drastically at the exceptional point $\abs{\gamma}=\abs{\thop}$.
In order to detect physics of the exceptional point, however, one would have to connect the isolated \PT-symmetric system to a Hermitian probe.
This motivates us to analyze an open \PT-symmetric system shown in Fig.~\ref{schema}, in which a non-Hermitian model is connected to Hermitian leads.
As we stressed above, we here make  an  observation for the transmission probability that is quite opposite to the one for the system~\eqref{H_PT2}.
The transmission probability exhibits strong signatures of non-Hermiticity in the weakly non-Hermitian regime, while it converges to a familiar Fabry-Perot-type peak structure.

We first present in Sec.~\ref{sec:single} a tutorial case of a single onsite non-Hermitian scatterer in an infinite tight-binding chain, in which we demonstrate that the transmission probability indicates the location of the exceptional point, a boundary between the \PT-unbroken phase and the \PT-broken phase.
Using the results of the transmission and reflection coefficients for the single scatterer, we obtain in Subsec.~\ref{subsec:FP} those for a \PT-symmetric pair of scatterers in the framework of the Fabry-Perot-type calculation.
We confirm these results in a more universal framework in Subsec.~\ref{subsec:M_L}.
We then describe in Sec.~\ref{sec:TandR} the peak structures of the transmission and reflection coefficients in terms of the Fabry-Perot resonance and point-spectral complex eigenvalues of resonant states.
We present in Sec.~\ref{sec:cont} the corresponding results for the continuum model with a pair of \PT-symmetric delta potentials, before concluding in Sec.~\ref{sec:conclusion}.
We present a brief review in App.~\ref{app:Siegert} for the point-spectral complex eigenvalues in open quantum systems and details of analytic calculations in Subsec.~\ref{subsec:M_L} in App.~\ref{app:ML}.

\section{Tutorial case of the single onsite scatterer $i\gamma$:
colliding peaks at the exceptional point}
\label{sec:single}

One of our motivations here is to see in an infinite open quantum system the physics of exceptional point typical in the two-site \PT-symmetric model~\eqref{H_PT2}.
It indeed manifests itself in the following 
simplest non-Hermitian scattering problem.
We here show its solutions for tutorial purposes;
we also utilize them in Subsecs.~\ref{subsec:FP} in solving the scattering problem of the \PT-symmetric scatterer by the Fabry-Perot-type formulation.

Let us consider the following minimal but non-Hermitian scattering problem:
\begin{align}
H:=\thop\sum_{x=-\infty}^\infty \qty( 
\dyad{ x+1}{ x} + \dyad{x+1 }{ x} 
) 
+ i\gamma \dyad{0},
\label{H_tb}
\end{align}
where $\thop$ denotes the amplitude of the hopping element, whose sign we do not specify for the moment.
The integer $x$ specifies a lattice site $x=0,\pm 1, \pm 2, \ldots$;
the lattice constant $a$ is chosen to be unity.
Our scattering potential is an isolated single onsite scatterer $i\gamma$ at $x=0$, where $\gamma$ is a real parameter with its sign not specified for the moment.
On each side of the scattering potential at the origin, the wave function is presumed to have the form:
\begin{align}
\psi_x = 
\begin{cases}
A e^{ikx} + B e^{-ikx}
& \mbox{for $x\le 0$},\\
 C e^{ikx}
& \mbox{for $x\ge 0$,}
\end{cases}
\label{inc-trm1}
\end{align}
where $\psi_x$ is short for $\braket{x}{\psi}$ and $k>0$.
Then, the continuity of the wave function at $x=0$ reinforces
\begin{align}
\psi_0=A+B=C.
\label{psi_0}
\end{align}

Under the boundary condition (\ref{inc-trm1}),
we solve the Schr\"odinger equation:
\begin{align}
H |\psi\rangle = E |\psi\rangle,
\label{eigen}
\end{align}
where
\begin{align}
|\psi\rangle:=(\cdots,\psi_{-1},\psi_0,\psi_1,\cdots)^t.
\end{align}
Away from the scattering potential $x=0$,
the Schr\"odinger equation gives the dispersion relation
\begin{align}
E(k):=2\thop \cos k.
\label{disp}
\end{align}
At $x=0$, on the other hand, it reads
\begin{align}
\thop\qty(\psi_{-1}+\psi_1)+i\gamma\psi_0=E(k)\psi_0.
\label{j=0}
\end{align}
Together with Eqs.~\eqref{inc-trm1}, \eqref{psi_0} and~\eqref{disp},
Eq.~\eqref{j=0} gives the transmission probability $T(k)$
of the incident wave in the form of
$T(k):=\abs{\calT(k)}^2$, where the transmission coefficient $\calT(k)$ is given by
\begin{align}
\calT(k)={C\over A}={2\thop\sin k\over 2\thop\sin k + \gamma}.
\label{C/A}
\end{align}
Remarkably, the scattering amplitude $\calT(k)$ diverges at
\begin{align}
k= -\arcsin{\gamma\over 2\thop},
\label{cond_div1}
\end{align}
which can occur when $\abs{\gamma}<2\abs{\thop}$.
The reflection probability is $R(k):=\abs{\calR(k)}^2$, where the reflection coefficient $\calR(k)$ is given by
\begin{align}
\calR(k):=\frac{B}{A}=\frac{-\gamma}{2\thop\sin k+\gamma},
\label{B/A}
\end{align}
and hence it has the same peak structure.
Physically, 
this divergence in the transmission and reflection coefficients can be regarded as an electronic analogue
of lasing,\cite{garmon,cpa1}
and is
due to a resonance pole incident on the real $k$ axis
(cf. discussion below on discrete eigenvalues under the Siegert boundary condition).
In Ref. \onlinecite{garmon}
this phenomenon is referred to as 
{\it resonance state in continuum} (RIC),
while in the literature
a similar resonance structure
has been also called a {\it spectral singularity}~\cite{mostaf_PRL,mostaf_PRA1,mostaf_PRA2,longhi_PRB,longhi_PRA,bender69,1986}.

For the gain $\gamma>0$
the divergence condition~\eqref{cond_div1} is met with
$k>0$, which is
a standard assumption in the scattering problem~\eqref{inc-trm1}.
This divergence realizes a situation of electronic analogue of lasing,
For a lossy potential $\gamma<0$, on the other hand,
the same condition is met with
$k<0$.
In this case, the divergence realizes a situation of coherent perfect absorption~\cite{cpa1,cpa2,cpa3}.

Note that
in normal crystals the hopping amplitude is usually negative: $\thop<0$.
However,
if we prepare a specific type of crystal, either electronic or photonic, in which $\thop>0$,
the role of gain and lossy potential is reversed.
Under this consideration, we hereafter fix $\gamma>0$ and $\thop<0$ for simplicity.

\begin{figure}
\centering
\includegraphics[width=70mm]{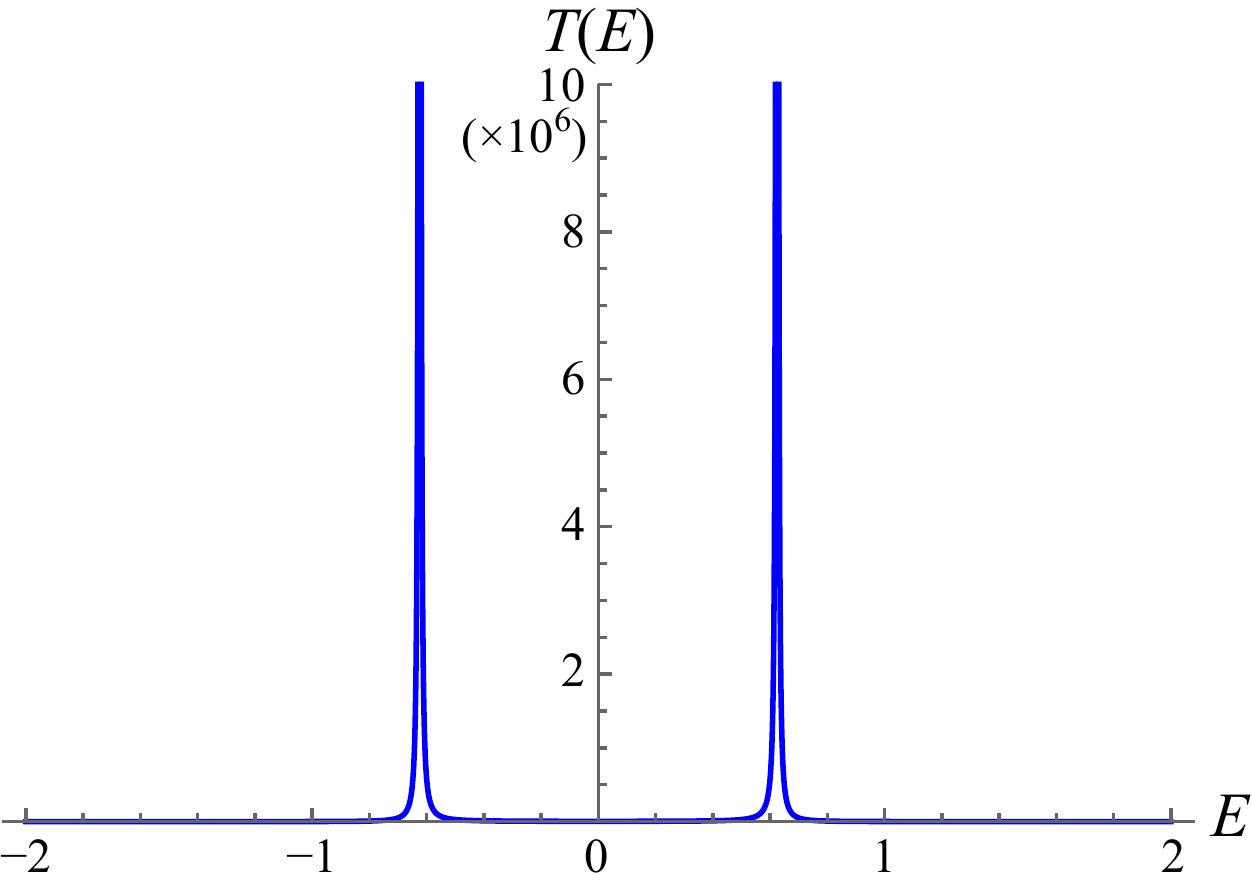}
\vspace{-\baselineskip}
\flushleft{(a)}\\
\vspace{\baselineskip}
\centering
\includegraphics[width=70mm]{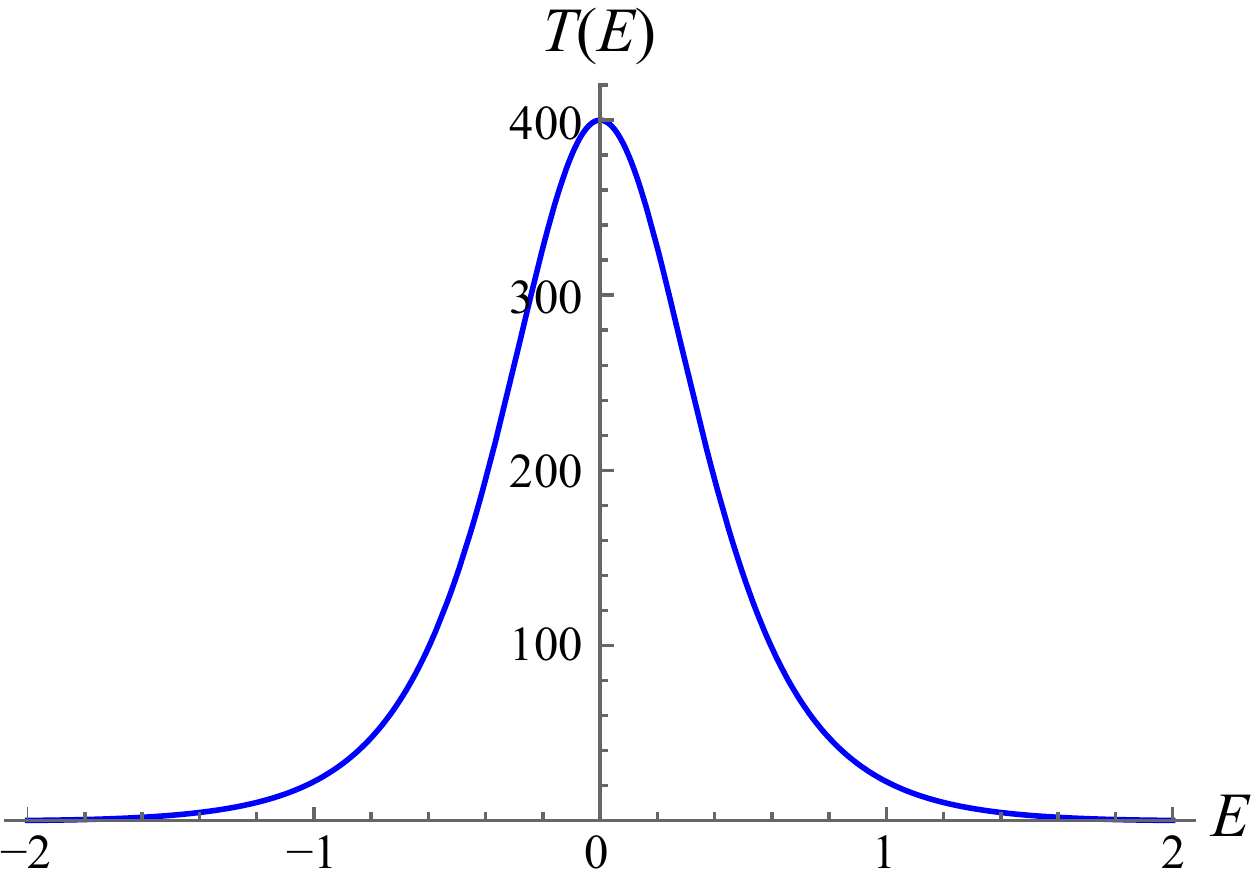}
\vspace{-\baselineskip}
\flushleft{(b)}\\
\vspace{\baselineskip}
\centering
\includegraphics[width=70mm]{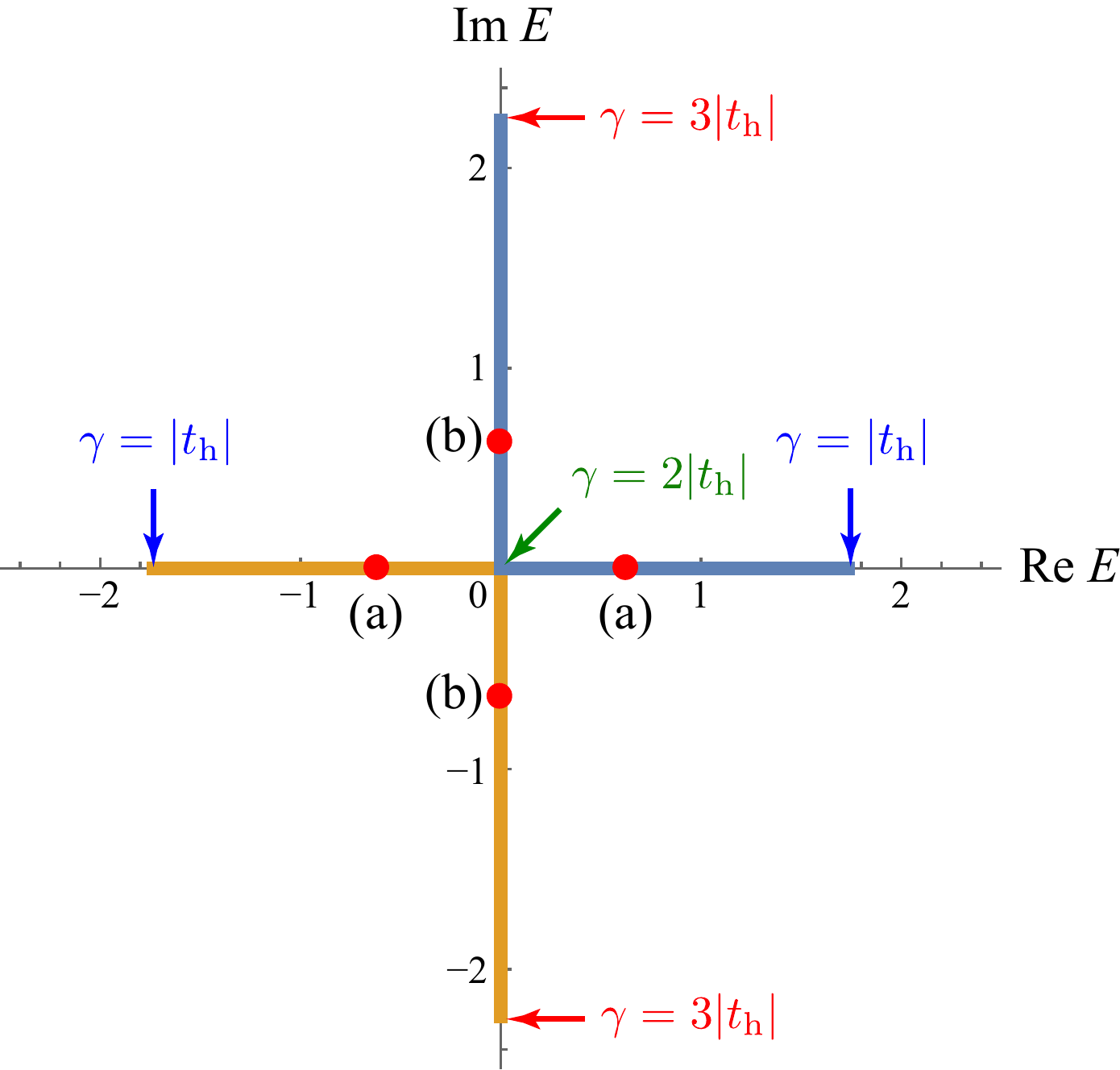}
\vspace{-\baselineskip}
\flushleft{(c)}\\
\caption{Scattering peaks in the complex potential model
(a) before and (b) after the collision of peaks at the exceptional point;
specifically, (a) $\gamma=1.9$ and (b) $\gamma=2.1$ both with $\thop=-1$.
(c) The change of the discrete eigenvalue~\eqref{EP1} (thick lines) in the complex energy plane due to the variation of $\gamma$ from $\abs{\thop}$ to $3\abs{\thop}$ with $\thop=-1$. The two complex eigenvalues collide at $E=0$ in the case of the exceptional point $\gamma=2\abs{\thop}$.
The dots on the real and imaginary axis respectively corresponds to the cases of (a) and (b).}
\label{fig:single}
\end{figure}
In Fig.~\ref{fig:single}, we plot the transmission probability $T(E)$ as a function of
energy $E$ in the two representative regimes:
(a) $\gamma<2\abs{\thop}$, and
(b) $\gamma>2\abs{\thop}$.
In panel (a)
[$\gamma=1.9$ with $\thop=-1$]
the transmission probability $T(E)$ 
shows divergent peaks at the two values of $E$ ($=E_1,E_2$) 
that satisfy Eq.~\eqref{cond_div1}.
The existence of such divergent peaks in $T(E)$ 
in this relatively weak $\gamma$ regime ($\abs{\gamma}<2\abs{\thop}$)
is a strongly non-Hermitian behavior atypical in Hermitian systems.
The two peaks at
$E_1$ and $E_2$
get closer to one another as $\gamma$ approaches $2\abs{\thop}$,
they ``collide'' at  $\gamma=2\abs{\thop}$, 
and transform into a broad peak after the collision
as it is in panel (b) [$\gamma=2.1$ ($\thop=-1$)].
This type of
a broad peak of height $T(E)>1$
in the strong $\gamma$ regime ($\gamma>2\abs{\thop}$)
is still different from a Hermitian behavior,
but less singular than the one
in the weak $\gamma$ regime.

This qualitative change of the behavior in $T(E)$ 
at $\gamma=2\abs{\thop}$
is related to the fact that this point falls on an exceptional point
in the parameter space.
To clarify this,
let us consider
discrete eigenvalues, namely point spectra, of an open system under the Siegert boundary condition~\cite{sieg,sieg1}, that is, we set $A=0$ in Eq.~\eqref{inc-trm1} so that there may be no incident wave.
This boundary condition is known to produce all discrete eigenvalues of open quantum systems, including bound states and resonant states, which coincides with all the poles of the $S$-matrix in general and of the transmission probability in one dimension~\cite{hatanoFDP};
see App.~\ref{app:Siegert}.

In the present case, the Siegert boundary condition signifies
\begin{align}
\psi_{-1}=\psi_1=\psi_0 e^{ik}.
\end{align}
Substituting this 
into Eq.~\eqref{j=0}, one can rewrite it as
\begin{align}
\qty(\thop\beta^2+i\gamma\beta-\thop)\psi_0=0,
\end{align}
where
$\beta=e^{ik}$.
In order to have a non-trivial solution $\psi_0\neq 0$, we solve the quadratic equation
\begin{align}
\thop\beta^2+i\gamma\beta-\thop=0,
\label{quad}
\end{align}
finding the solutions $\beta_\pm=\qty(-i\gamma\pm\sqrt{4\thop^2-\gamma^2})/(2\thop)$, which produce 
the two eigenvalues in the complex energy plane in the form
\begin{align}
E_\pm&=\thop\qty(\beta_\pm+\frac{1}{\beta_\pm})
= \pm \sqrt{4\thop^2-\gamma^2}.
\label{EP1}
\end{align}
Indeed, the divergence condition~\eqref{cond_div1} is equivalent to Eq.~\eqref{quad}. The transmission probability $T=|C/A|^2$  becomes infinite
at its divergence implying $A\rightarrow 0$, while under the Siegert boundary condition, the condition $A=0$ is preassigned.

Figure~\ref{fig:single}(c) shows how the two eigenvalues change in the complex energy plane as we vary the parameter $\gamma/\abs{\thop}$.
When $\gamma<2\abs{\thop}$,
Eq.~\eqref{quad} has two solutions $\beta_+$ and $\beta_-$ with their corresponding eigen-wave-number $k_\pm = -i\log\beta_\pm$
are both on the real axis.
The eigenvalues $E_\pm = 2t\cos k_\pm$
are also on the real axis.
Correspondingly, the transmission probability $T(k)$ and $T(E)$
diverge at these values of $k$ or $E$.
When $\gamma>2\abs{\thop}$,
the two solutions 
$E_\pm$ become purely imaginary.
Therefore,
$T(E)$
do not show any longer a divergence on the real axis
on which they are defined.
The transmission probability $T(E)$ instead shows a typical Breit-Wigner peak at $E=0$.

The condition $\gamma=2\abs{\thop}$ sets an exceptional point
at which the two real eigenvalues collide and turn
into a pair of two purely imaginary values, as we can see in Eq.~\eqref{EP1}.
The collision occurs at $E=0$ in the complex $E$-plane.
In terms of $k$,
the story is almost parallel except that the collision occurs
at $k=\pi/2$.

Let us finally comment on the unitarity of the $S$-matrix, 
\textit{i.e.}, on the behavior of the sum of the transmission and reflection probabilities.
Using Eqs.~\eqref{C/A} and~\eqref{B/A}, we have
\begin{align}
T(k)+R(k)=\frac{4\thop^2\sin^2 k+\gamma^2}%
{4\thop^2\sin^2k+\gamma^2+4\thop\gamma\sin k}.
\end{align}
In the standard case of $\thop<0$, the sign of the last term in the denominator depends on the sign of the parameter $\gamma$ because $k>0$.
Therefore, $T(k)+R(k)>1$ if $\gamma>0$ and $T(k)+R(k)<1$ if $\gamma<0$.
This is indeed consistent with the fact that the scattering potential is a source if $\gamma>0$ and a sink if $\gamma<0$.
Anyway, the unitarity of the $S$-matrix is always broken: $T+R\neq 1$.

\section{Scattering problem of a \PT-symmetric scatterer}
\label{sec:PT}

Let us now come to the main point of the present paper and consider the scattering problem of a pair of on-site scatterers
$i\gamma$ and $-i\gamma$, as is shown in Fig.~\ref{schema}.
The model is specified by the Hamiltonian
\begin{align}
H&=
\thop\sum_{x=-\infty}^\infty 
\qty( 
\dyad{x+1}{x} + \dyad{x+1}{x}
) 
\nonumber \\
&+ 
V_0\dyad{0}+V_L\dyad{L},
\label{H_PT}
\end{align}
where $V_0$ and $V_L$ are on-site scattering potentials,
for which, unless otherwise specified,
we consider the non-Hermitian \PT-symmetric case:
$V_0= i\gamma$ and $V_L= -i\gamma$.
(We mention other cases at the end of Subsec.~\ref{subsec-eig}.)
We also consider the case $\gamma>0$.
The scattering region falls on $x\in [0,L]$.
The case of $L=2$ has been studied in Ref.~\onlinecite{garmon}.
We can confirm the \PT-symmetry by reflecting the system with respect to the point $x=L/2$ and take the complex conjugation.
The sign of $\thop$ depends on the actual physical setups, as is in the previous section~\ref{sec:single}, but we take $\thop<0$ except where indicated.

We will find the transmission and reflection probabilities in two ways.
We first use a Fabry-Perot-type formulation in Subsec.~\ref{subsec:FP}.
We then derive the same formulas by using a more universal formulation of solving the scattering problem in Subsec.~\ref{subsec:M_L}.

\subsection{Fabry-Perot formulas for the transmission and reflection probabilities}
\label{subsec:FP}

We first assume the incident and reflective waves on the left of the scattering region, while the transmissive wave on its right:
\begin{align}
\psi_x = 
\begin{cases}
A e^{ikx} + B e^{-ikx}
& \mbox{for $x\le 0$},\\
 C e^{ikx}
& \mbox{for $x\ge L$}.
\end{cases}
\label{inc-trm}
\end{align}
Our aim is to obtain the transmission and reflection coefficients, $\calT:=C/A$ and $\calR:=B/A$.

\begin{figure}
\centering
\includegraphics[width=70mm]{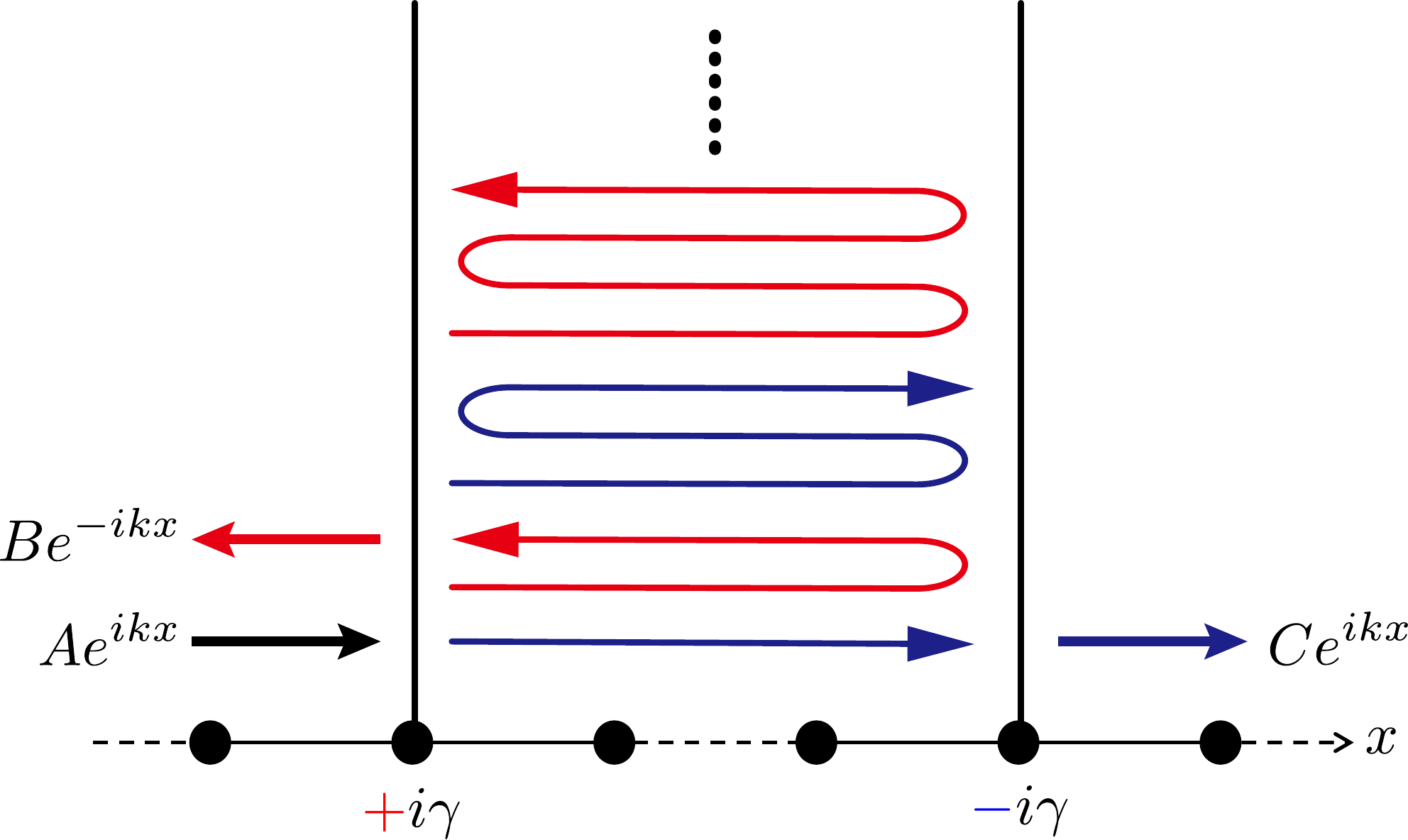}
\caption{The Fabry-Perot-type calculation of the contributions to the transmissive wave (the blue lines) and to the reflective wave (the red lines).}
\label{fig:FP}
\end{figure}

We try to find the transmissive wave from the left of the potential $+i\gamma$ to the right of the potential $-i\gamma$ as a superposition of
\begin{enumerate}
\renewcommand{\labelenumi}{(\roman{enumi})}
\item the wave that transmits each of the two potentials,
\item the wave that transmits the left potential, reflects at the right potential, reflects back at the left potential and transmits the right potential,
\item the wave that goes back and forth between the two potentials twice more,
\item and so on;
\end{enumerate}
see Fig.~\ref{fig:FP}.
More specifically, for an incident wave of amplitude $A$ at $x=0$, we represent the transmissive wave $\psi_L:=\braket{L}{\psi}=C e^{ikL}$ in terms of the following infinite series:
\begin{align}
Ce^{ikL}&=A \calT_0 e^{ikL} \calT_L
+ A \calT_0 e^{ikL} \calR_L e^{ikL}\tilde{\calR}_0 e^{ikL} \calT_L+ \cdots
\nonumber \\
&= {A \calT_0 \calT_L e^{ikL} 
\over 1- \calR_L \tilde{\calR}_0 e^{2ikL}},
\label{FPser}
\end{align}
where
$\calT_0$, $\calT_L$, $\calR_L$ and $\tilde{\calR}_0$ are elements of the $S$-matrices for the potentials $\pm i\gamma$ at $x=0$ and $x=L$; namely, the $S$-matrix for the potential $+i\gamma$ at $x=0$ reads
\begin{align}
S_0=
\mqty(
\calR_0 & \tilde{\calT}_0 \\
\calT_0 & \tilde{\calR}_0
),
\label{S0}
\end{align}
while that for the potential $-i\gamma$ at $x=L$ reads
\begin{align}
S_L=\mqty(
\calR_L &  \tilde{\calT}_L \\
\calT_L  &  \tilde{\calR}_L  
).
\label{SL}
\end{align}

We know from the results in Sec.~\ref{sec:single} that
\begin{align}
\calT_0&=\frac{2\thop\sin k}{2\thop\sin k+\gamma},
\qquad
\calR_0=\frac{-\gamma}{2\thop\sin k +\gamma}.
\label{TR_0}
\end{align}
Since the scattering problem for a single on-site scatterer is symmetric with respect to the potential, the coefficients due to the incident wave from the right should be equal to the equivalent coefficients from the left:
$\tilde{\calT}_0 = \calT_0$ and $\tilde{\calR}_0 = \calR_0$.

For $\calT_L$ and $\calR_L$, we only need to flip the sign of the parameter $\gamma$ because the potential there is $-i\gamma$ instead of $+i\gamma$. Therefore, we have
\begin{align}
\calT_L&={2\thop\sin k\over 2\thop\sin k- \gamma},
\qquad
\calR_L={\gamma\over  2\thop\sin k-\gamma}.
\label{TR_L}
\end{align}
According to the same argument for $S_0$, we should have $\tilde{\calT}_L = \calT_L$ and $\tilde{\calR}_L = \calR_L$.

Substituting the expressions in Eqs.~\eqref{TR_0} and~\eqref{TR_L} into Eq.~\eqref{FPser}, we arrive at
\begin{align}
\calT(k)=\frac{C}{A}=\frac{4\thop^2\sin^2 k}%
{4\thop^2\sin^2 k+ \gamma^2(e^{2ikL}-1)};
\label{FP-T}
\end{align}
We will find the same expression from the standard way of solving the scattering problem in the next subsection~\ref{subsec:M_L}.

We can derive the reflection coefficient $\calR$ in the same way.
In parallel with Eq.~\eqref{FPser}, we find the amplitude $B$ for the reflective wave in the form
\begin{align}
B&=A \calR_0 + A \calT_0 e^{ikL} \calR_L e^{ikL} \tilde{\calT}_0
\nonumber\\
&+ A \calT_0 e^{ikL} \calR_L e^{ikL} \tilde{\calR}_0 
e^{ikL} \calR_L e^{ikL} \tilde{\calT}_0
+ \cdots
\nonumber \\
&= A \calR_0 + {A\calT_0 \calR_L \tilde{\calT}_0 e^{2ikL} 
\over 1- \calR_L \tilde{\calR}_0 e^{2ikL}}.
\label{BA_L}
\end{align}
After straightforward algebra, we find 
\begin{align}
\calR(k)&={B \over A}=\calR_0+\calR_L e^{2ikL}\calT(k)
\nonumber\\
&={\gamma(2\thop\sin k- \gamma)(e^{2ikL}-1)
\over 4\thop^2\sin^2 k+ \gamma^2(e^{2ikL}-1)}.
\label{FP-R}
\end{align}
We will also find the same expression in the next subsection~\ref{subsec:M_L}.

We can similarly find the transmission and reflection coefficients $\tilde{\calT}$ and $\tilde{\calR}$ due to the incident wave from the right.
These coefficients have the expressions in which the sign of the parameter $\gamma$ is flipped in $\calT$ and $\calR$ because the first potential that the incident wave from the right meets is the one of $-i\gamma$ instead of $+i\gamma$.
Since $\calT(k)$ is an even function of $\gamma$, we easily find $\tilde{\calT}(k)=\calT(k)$, but $\abs{\calR}\neq\abs{\tilde{\calR}}$ in contrast;
in other words,
our transmission coefficients are reciprocal, while
the reflection coefficients are non-reciprocal.
This type of non-reciprocal transport is possible in a non-Hermitian system with parity broken but transposition unbroken 
(for details, see Sec.~6.1 of Ref.~\onlinecite{rev1}).

\subsection{More universal way of solving the problem}
\label{subsec:M_L}

We can solve the potential-scattering problem in the following standard formulation too.
We again assume the form~\eqref{inc-trm}.
The Schr\"odinger equation at the sites $x=0,1,2,\cdots,L$ thereby read
\begin{align}
\thop \qty(\psi_{-1}+\psi_1)+i\gamma\psi_0 &= E \psi_0,
\nonumber \\
\thop \qty(\psi_{0}+\psi_2) &= E \psi_1.
\nonumber \\
\cdots&
\nonumber \\
\thop \qty(\psi_{L-2}+\psi_{L}) &= E \psi_{L-1}.
\nonumber \\
\thop \qty(\psi_{L-1}+\psi_{L+1})-i\gamma\psi_L &= E \psi_L.
\label{listA}
\end{align}
The remaining equations for $x<0$ and $x>L$ simply give the dispersion relation~\eqref{disp}.

We can cast the open set of equations~\eqref{listA} into a closed matrix equation in the following way.
In the list of equations~\eqref{listA}, we express
the wave-function amplitudes one step outside of the scattering region,
namely $\psi_{-1}$ and $\psi_{L+1}$, in terms of the amplitudes 
inside the scattering region $x\in [0,L]$.
Utilizing Eq.~\eqref{inc-trm}, we have
\begin{align}
\psi_{-1} &= A e^{-ik}+B e^{ik}
\nonumber \\
&= A e^{-ik}+(\psi_0-A) e^{ik}
\nonumber \\
&= -2i A \sin k+e^{ik}\psi_0 
\label{inc2}
\end{align}
and
\begin{align}
\psi_{L+1} = C e^{ik(L+1)} = e^{ik} \psi_{L}
\label{trm2}
\end{align}
because $\psi_0=A+B$ and $\psi_L=Ce^{ikL}$.
Inserting Eqs.~\eqref{inc2} and \eqref{trm2} into the first and last equations of Eqs.~\eqref{listA}, we arrive at the closed $(L+1)$-dimensional matrix equation
\begin{align}
M_L 
\mqty(
\psi_0\\
\psi_1\\
\vdots\\
\psi_{L-1}\\
\psi_{L}
)
=
\mqty(
\tilde{A}\\
0\\
\vdots\\
0\\
0
),
\label{ML1}
\end{align}
where
\begin{align}
\tilde{A}=2i A\thop  \sin k,
\label{tildeA}
\end{align}
and an $(L+1)\times (L+1)$ matrix
\begin{align}
M_L&=H_L-E(k)I_{L+1}
\nonumber\\
&=H_L-\thop\qty(e^{ik}+e^{-ik}) I_{L+1}
\label{ML2}
\end{align}
with $H_L$ denoting an effective Hamiltonian matrix
\begin{align}
H_L=\mqty(
i\gamma+\thop e^{ik}  &  \thop &&& \\
\thop &  & \thop && \\
& \thop &  &  \ddots & \\
 && \ddots &  &  \thop \\
&&& \thop  &  -i\gamma+\thop e^{ik} 
)
\label{HL1}
\end{align}
and $I_{L+1}$ denoting the $(L+1)$-dimensional identity matrix. 

Using Eqs.~\eqref{ML1}--\eqref{HL1}, we can derive the transmission and reflection coefficients as follows.
Equation~\eqref{ML1} implies
\begin{align}
\psi_{0}&=\left({M_L}^{-1}\right)_{1,1}\tilde{A},
\\
\psi_{L}&=\left({M_L}^{-1}\right)_{L+1,1}\tilde{A},
\end{align}
which read
\begin{align}
1+\frac{B}{A}&=2i \thop \left({M_L}^{-1}\right)_{1,1} \sin k,
\label{MI-B/A}
\\
\frac{C}{A}e^{ikL}&=2i \thop \left({M_L}^{-1}\right)_{L+1,1} \sin k.
\label{MI-C/A}
\end{align}
We can thus obtain $\calT=C/A$ and $\calR=B/A$ 
by finding the two elements of the inverted matrix ${M_L}^{-1}$.

The inversion of the matrix $M_L$ involves computation of matrix determinants by means of recursion equations. 
The algebra given in App.~\ref{app:ML} produces exactly the same expressions as Eqs.~\eqref{FP-T} and~\eqref{FP-R}.
We can thus validate the calculation based on the Fabry-Perot-type formulation.

\section{Contrasting behavior of $T(k)$ and $R(k)$  
in the weak and strong $\gamma$ regimes}
\label{sec:TandR}

\begin{figure}
\centering
\includegraphics[width=70mm]{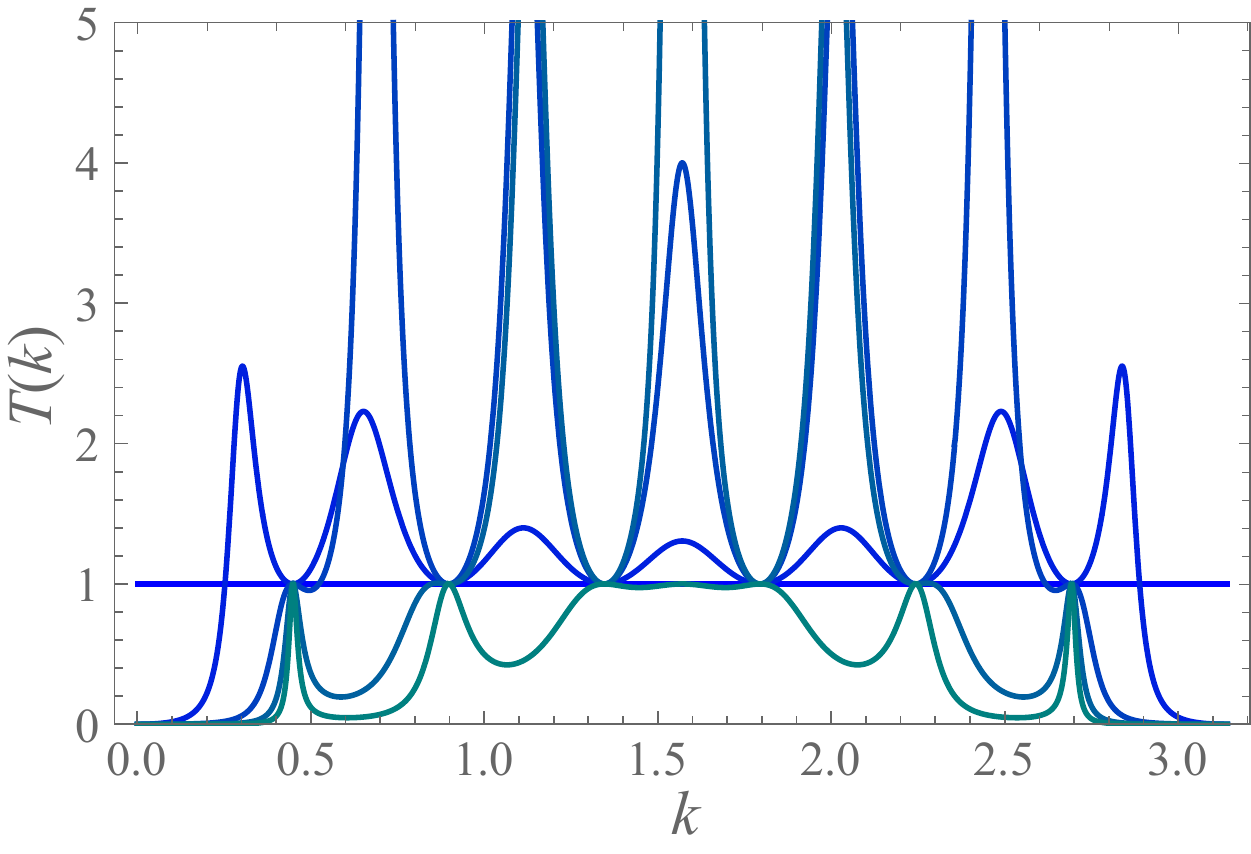}
\vspace{-\baselineskip}
\flushleft{(a)}\\
\vspace{\baselineskip}
\centering
\hspace{-3.5mm}
\includegraphics[width=73mm]{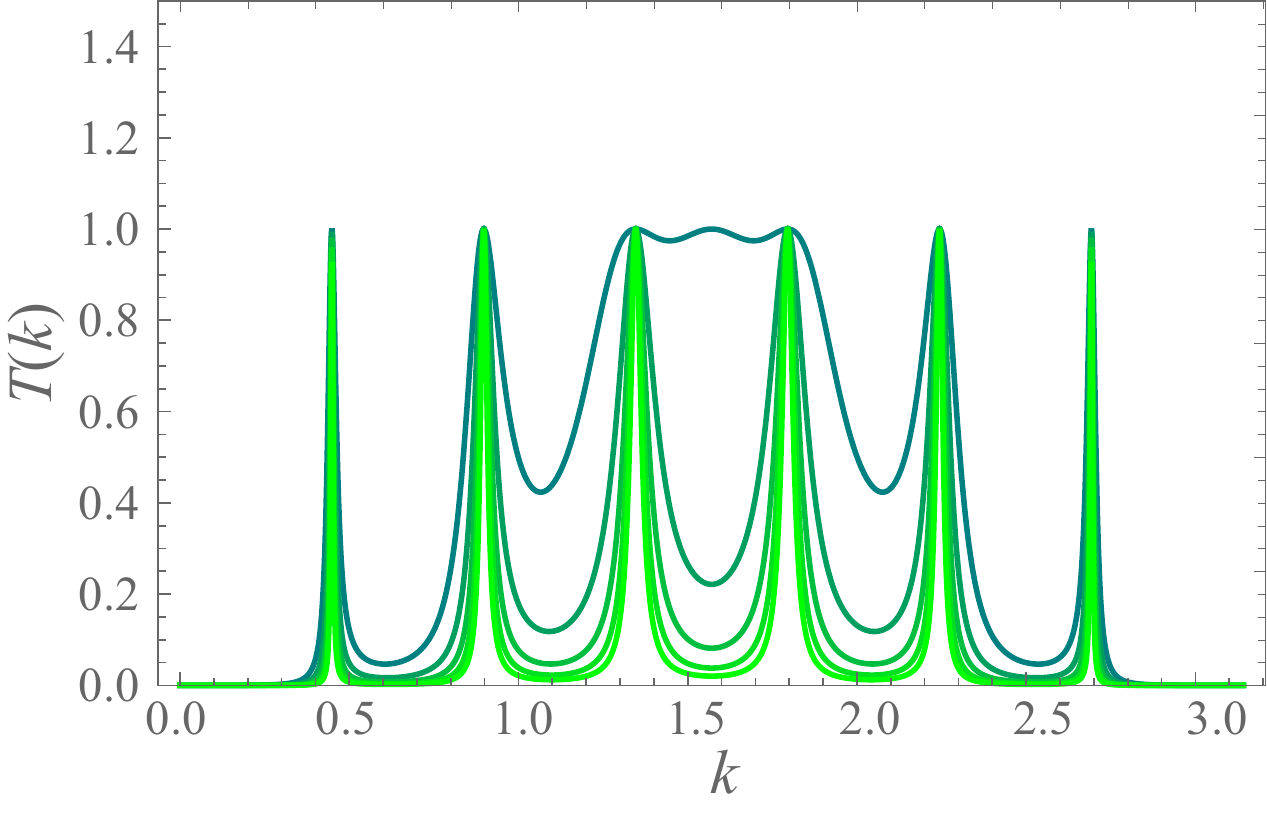}
\vspace{-\baselineskip}
\flushleft{(b)}
\caption{Variation of 
the transmission probability $T(k)$ for $L=7$; 
(a) in the weak regime $\gamma<2\abs{\thop}$ ($\gamma /\abs{\thop}=0, 0.5, 1, 1.5, 2$) and
(b) in the strong regime $\gamma>2\abs{\thop}$ ($\gamma /\abs{\thop}=2, 2.5, 3, 3.5, 4$).
The color varies from blue to green as we increase $\gamma$ from $0$ to $4\abs{\thop}$.
}
\label{T_L7}
\end{figure}

\subsection{Transmission and reflection probabilities}

Figure~\ref{T_L7} shows the transmission probability $T(k)=|C/A|^2$ of the system with $L=7$, which exemplifies the generic case of odd $L$.
The behavior is quite different between a region of weak non-Hermiticity, namely $\gamma<2\abs{\thop}$, and a region of strong non-Hermiticity, $\gamma>2\abs{\thop}$.
Note that the critical value $\gamma=2\abs{\thop}$ corresponds to the exceptional point in the problem of the single onsite scatterer that we considered in Sec.~\ref{sec:single}.

In the region of weak non-Hermiticity $\gamma<2\abs{\thop}$,
the transmission probability $T(k)$ (represented by bluish curves)
shows a peak higher than unity.
Such behavior is common in a non-Hermitian scattering problem; 
in the most prototypical case of a single isolated onsite scatterer $i\gamma$ analyzed in Sec.~\ref{sec:single}, the transmission probability $T(k)$ can not only exceed unity but can diverge for some values of $k$  at the point that satisfies Eq.~\eqref{cond_div1}.
In the present \PT-symmetric case, the transmission peaks are not generally divergent but exceed unity particularly in the region of weak non-Hermiticity.
The divergence occurs for specific values of $\gamma$ given in Eq.~\eqref{gamma_div} at the values of $k$ given in Eq.~\eqref{k_div}.

In the region $\gamma>2\abs{\thop}$, on the other hand, the situation is superficially Hermitian in contrast.
The transmission probability $T$ (represented by greenish curves) still show peaks but their height is bounded by unity, $T(k)\le 1$, which is common in the Hermitian case.
These peaks are consistent with the conventional Fabry-Perot-type resonances at the wave numbers
\begin{align}
k={n\pi \over L},
\label{FPcond}
\end{align}
where $n=1,2,\cdots,L-1$.
In Subsec.~\ref{subsec:peakT}, we describe these results in terms of the analytical formula of the transmission probability.

\begin{figure}
\centering
\includegraphics[width=70mm]{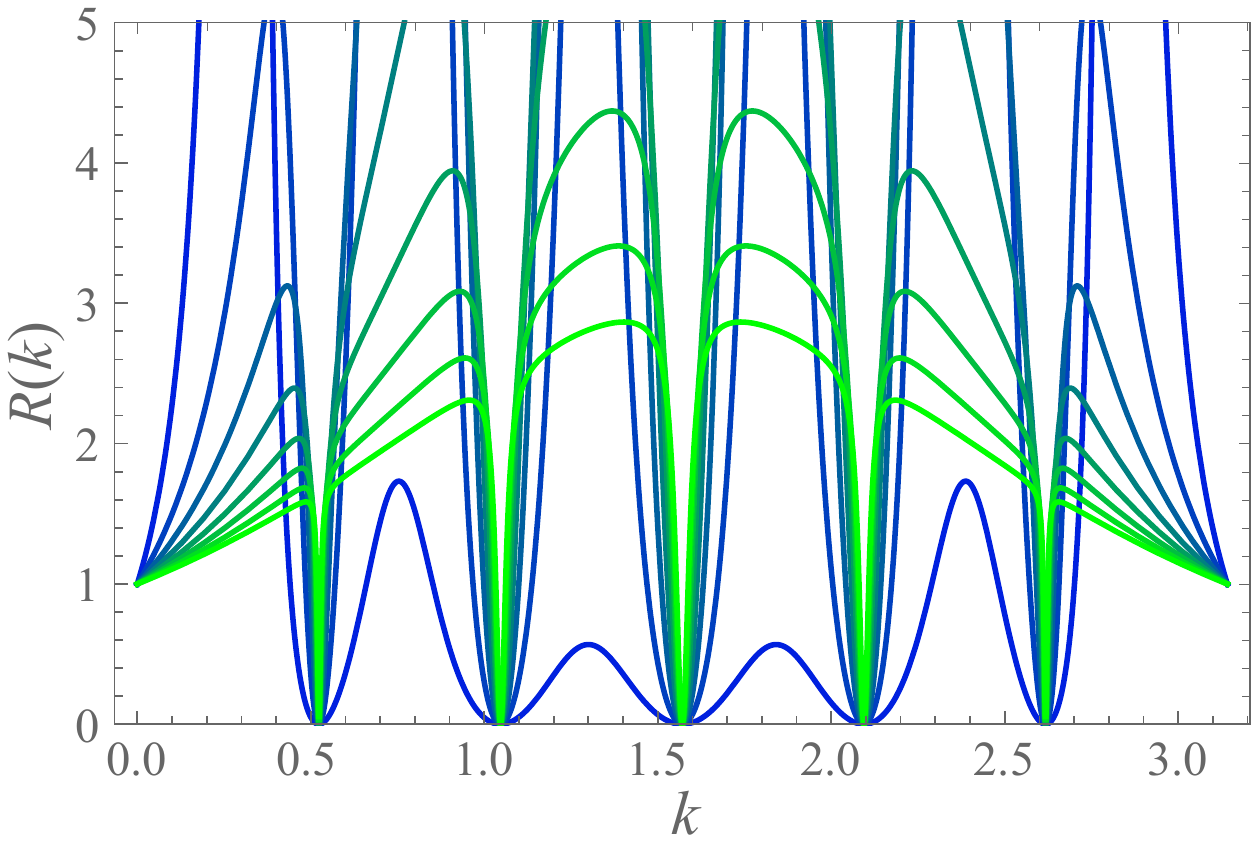}
\vspace{-\baselineskip}
\flushleft{(a)}\\
\vspace{\baselineskip}
\centering
\includegraphics[width=70mm]{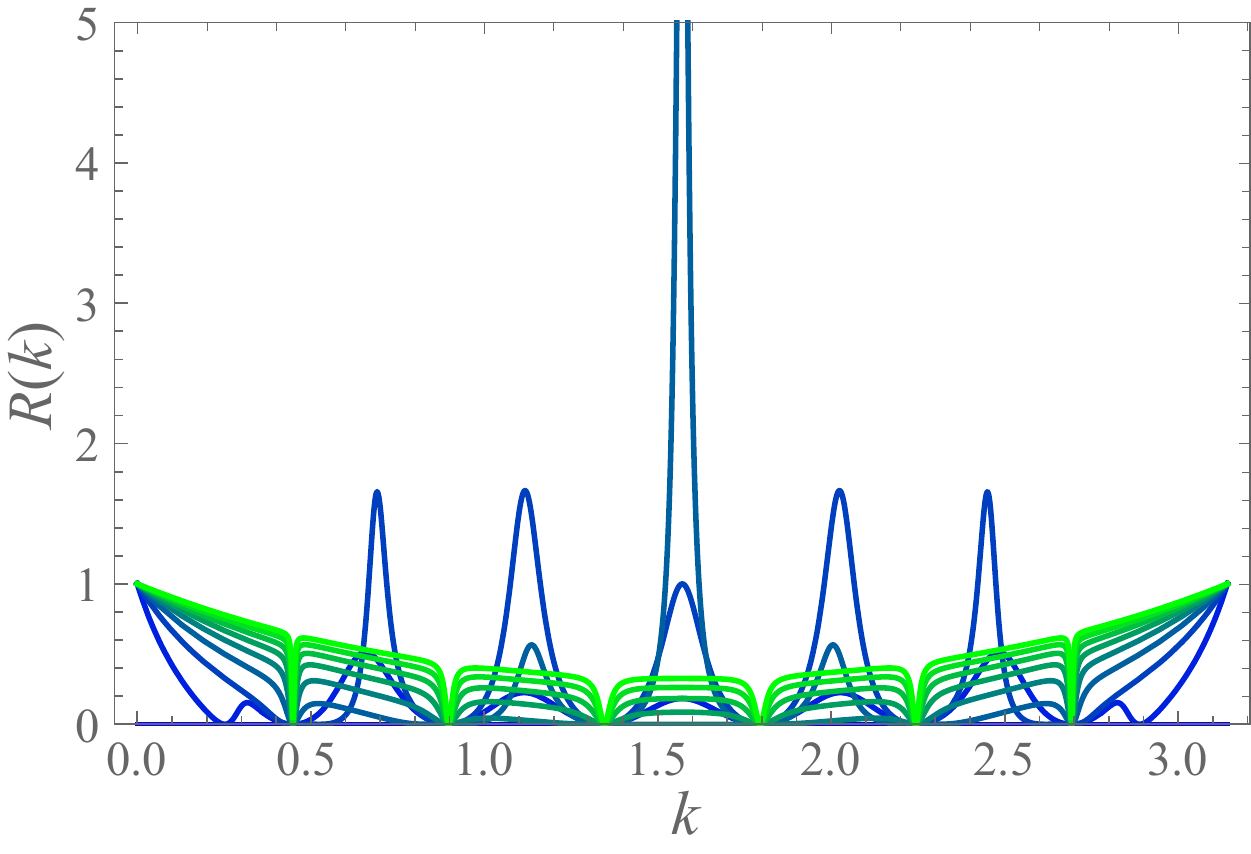}
\vspace{-\baselineskip}
\flushleft{(b)}
\caption{
Variation of the reflection probabilities for $L=7$ 
at $\gamma /\abs{\thop}=0.5, 1, 1.5,\cdots, 3.5, 4$; 
(a) $R(k)$ for the incident wave from the left and
(b) $\tilde{R}(k)$ for the  incident wave from the right.
}
\label{R_L7}
\end{figure}

The behavior of the reflection probability $R(k)$, on the other hand, is different from the one common in the Hermitian Fabry-Perot case;
In Fig.~\ref{R_L7}, $R(k)$ exhibits dips at the wave numbers given in Eq.~\eqref{FPcond}, which itself is consistent with the Fabry-Perot-type resonance, 
but its magnitude elsewhere breaks the unitarity of the $S$-matrix,
\begin{align}
T+R = |\calT|^2+|\calR|^2 \neq 1,
\label{neq1}
\end{align}
even in the superficially Hermitian regime $\gamma>2\abs{\thop}$.
The reflection probability due to the incident wave from the left exceeds unity except for the dips, while that due to the incident wave from the right is suppressed to less than unity.
We present in Subsec.~\ref{subsec:peakR} approximate functions of these profiles.

\subsection{Understanding the peak structure of the transmission probability}
\label{subsec:peakT}

We here analyze the drastic change of $T(k)$ at $\gamma=2\abs{\thop}$ that we found in Fig.~\ref{T_L7}, from the point of view of the analytic formula~\eqref{FP-T} for the transmission coefficient.
Let us first note that
Eq.~\eqref{FP-T} yields the transmission probability $T(k):=\abs{\calT(k)}^2$ in the form
\begin{align}
T(k)=\frac{4\thop^4\sin^4 k}%
{4\thop^4\sin^4 k + \gamma^2 (\gamma^2 - 4\thop^2\sin^2 k)\sin^2 kL},
\label{Tk}
\end{align}
which is greater than unity, $T(k)>1$, for $\gamma<2\abs{\thop}$ in the following range of $k$:
\begin{align}
\arcsin{\gamma\over 2\abs{\thop}} <k < \pi -\arcsin{\gamma\over 2\abs{\thop}}
\label{range_k}
\end{align}
so that the factor $(\gamma^2-4\thop^2\sin^2 k)$ may be positive.
In the Fabry-Perot regime $\gamma>2\abs{\thop}$, on the other hand, it is always less than unity:
\begin{align}
T(k)<1\quad\mbox{for all $k$ when $\gamma>2\abs{\thop}$}.
\end{align}

For $T(k)$ to be divergent, the denominator of Eq.~\eqref{FP-T} must vanish.
Since the first term $4\thop^2\sin^2 k$ of the denominator is real and positive except at $k=0$ and $k=\pi$, a divergent peak can appear for a generic value of $k$ only when the complex factor $e^{2ikL}$ happens to be real; in other words,
\begin{align}
e^{2ikL}=\pm 1.
\end{align}

In the cases of $e^{2ikL}=1$, peaks emerge, but do not diverge, because the second term of the denominator of Eq.~\eqref{FP-T} vanishes, and one trivially finds
$\calT(k)=1$.
This indeed corresponds to the spuriously Hermitian Fabry-Perot peaks that we observe in the regime of $\gamma>2\abs{\thop}$ (Fig.~\ref{T_L7}(b)); indeed, $e^{2ikL}=1$ is the Fabry-Perot resonance condition.
When $\gamma>2\abs{\thop}$, we have the perfect transmission $T(k)=1$ at each resonance point while $T(k)<1$ off resonance because the second term dominates the denominator in Eq.~\eqref{FP-T} if $\gamma>2\abs{\thop}$ and $e^{ikL}\ne 1$.

Hereafter throughout the present subsection~\ref{subsec:peakT}, we discuss the cases of $e^{2ikL}=-1$.
In such cases, the denominator of Eq.~\eqref{FP-T} vanishes if
\begin{align}
\gamma=\sqrt{2} \abs{\thop}\sin k.
\label{gamma_div}
\end{align}
This corresponds to the divergent peaks of $T(k)$ in the regime of $\gamma<2\abs{\thop}$ (Fig.~\ref{T_L7}(a)).
To be more explicit, $e^{2ikL}=-1$ is satisfied at the values of $k$
such that
\begin{align}
k=k_n:={(2n-1)\pi\over 2L},
\label{k_div}
\end{align}
where
$n=1,2,\cdots,L$.
The transmission probability $T(k)$ is divergent at $k=k_n$
when we tune the value of $\gamma$ close to $\gamma_n:=\sqrt{2} \abs{\thop}\sin k_n$, which is possible only when $\gamma\le \sqrt{2}\abs{\thop}<2\abs{\thop}$.
The integer $n$ that is responsible for the divergence changes as we vary $\gamma$.
When we turn on $\gamma$, the first pair of peaks develop at $k=k_1$ and $k=k_L$, and diverge when we tune $\gamma$ up to $\gamma_1$.
As we increase $\gamma$ from $\gamma_1$ towards $\gamma_2$, the first pair of peaks subside and the second pair of peaks start to grow at $k_2$ and $k_{L-1}$, which diverge when $\gamma$ reaches $\gamma_2$. 
As we further increase $\gamma$, successive peaks appear at $k=k_3,\ldots$ on the side of $k<\pi/2$ and at $k=k_{L-2},\ldots$ on the side of $k>\pi/2$.
The positions of the divergent peaks that diverge thus moves toward $k=\pi/2$ from both sides; see the video 1 in the supplementary material.

When $\gamma$ goes to $\sqrt{2} \abs{\thop}$ from below, what happens depends on the parity of $L$.
If $L$ is odd with $L=2l-1$ as in Fig.~\ref{T_L7}(a), the last divergent peak appears at the mid-point $k=k_l=\pi/2$ because it is contained in the set \eqref{k_div}.
At this point, Eq.~\eqref{FP-T} reduces to 
\begin{align}
\calT(\pi/2)=
{2\thop^2\over 2\thop^2 - \gamma^2},
\label{pi/2}
\end{align}
which diverges at $\gamma = \sqrt{2} \abs{\thop}$.
Beyond this value of $\gamma$, the transmission probability at the mid-point subsides, and eventually lowers the unity when $\gamma>2\abs{\thop}$ (Fig.~\ref{T_L7}(b)).
Since the peak at $k=\pi/2$ is the last divergent one as we increase $\gamma$, other peaks decay at even smaller values of $\gamma$, and hence no peak exceeds the unity beyond $\gamma = 2 \abs{\thop}$.
This explains why the transmission probability $T(k)$ becomes superficially Hermitian in the regime of $\gamma>2\abs{\thop}$.
It may deserve to mention that $T(\pi/2)$ turns from a maximum to a minimum at $\gamma=2\abs{\thop}$; see Fig. \ref{T_L7}(a) and (b).

\begin{figure}
\centering
\includegraphics[width=70mm]{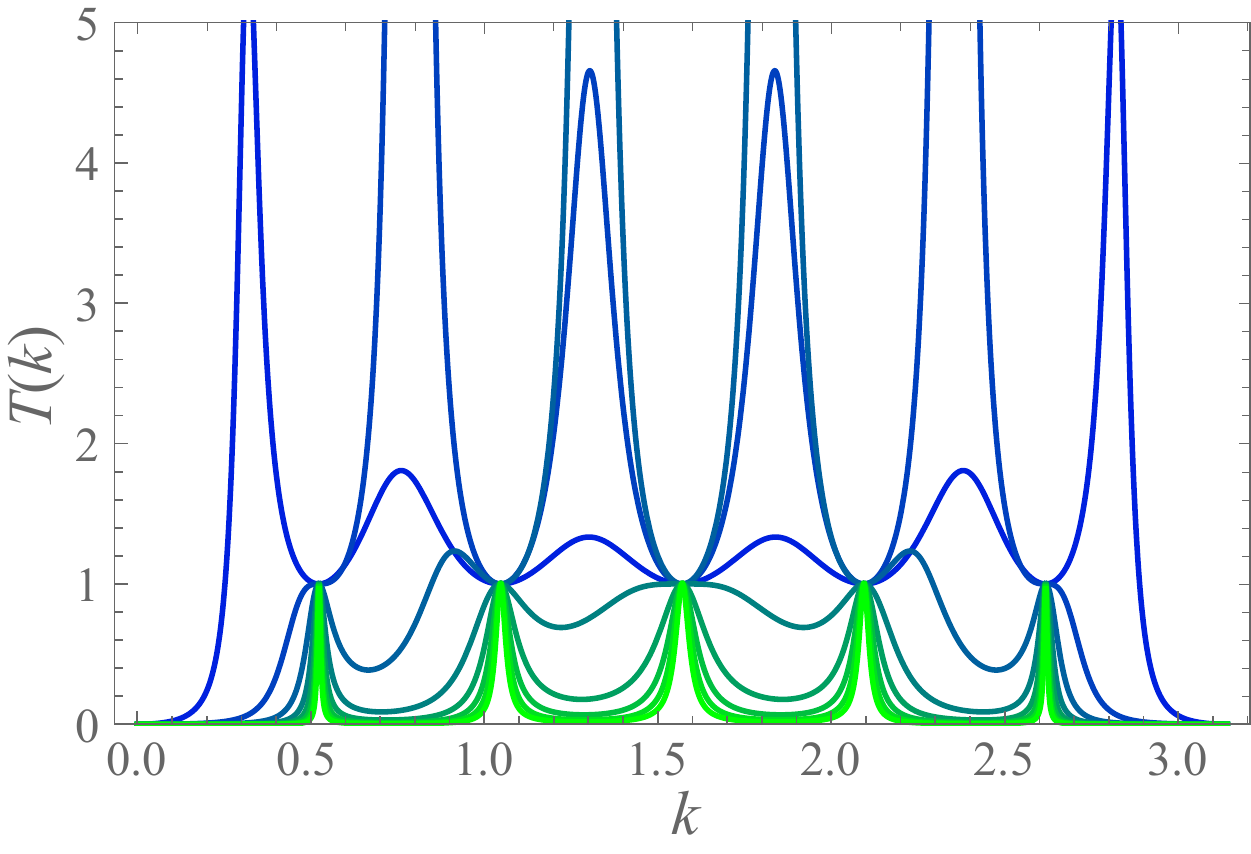}
\caption{Variation of 
the transmission probability $T(k)$ for $L=6$
at $\gamma /\abs{\thop}=0.5, 1, 1.5,\cdots, 3.5, 4$.
The color varies from blue to green as we increase $\gamma$ from $0$ to $4\abs{\thop}$.
The bluish curves correspond to the weak regime $\gamma<2\abs{\thop}$, while the greenish ones to strong regime $\gamma>2\abs{\thop}$.
}
\label{T_L6}
\end{figure}

If $L$ is even with $L=2l$ as in Fig.~\ref{T_L6}, on the other hand, 
the last divergence occurs not at $k=\pi/2$ but at $k_l$ and $k_{l+1}$, \textit{i.e.}, at
\begin{align}
k={\pi\over 2}\pm {\pi\over 2L}
\end{align}
at the same time.
At these values of $k$, Eq.~\eqref{FP-T} reduces to
\begin{align}
\calT(k)=
\frac{\thop^2\qty[1+ \cos \qty(\pi/ L)]}%
{\thop^2\qty[1+ \cos \qty(\pi/ L)] - \gamma^2}.
\label{last2}
\end{align}
Therefore the divergence of the last peaks at $k=k_l$ and $k=k_{l+1}$ occurs when we increase $\gamma$ up to
\begin{align}
\gamma = \abs{\thop} \sqrt{1+ \cos \frac{\pi}{L}}<\sqrt{2}\abs{\thop}.
\end{align}
When we further increase so that
\begin{align}
\gamma > \sqrt{2}\abs{\thop} \sqrt{1+ \cos {\pi\over L}}>\sqrt{2}\abs{\thop},
\label{last2cond}
\end{align}
Eq.~\eqref{last2} gives
$T(k_l)=T(k_{l+1})\le 1$, \textit{i.e.}, a superficially Hermitian result.
Meanwhile, $T(\pi/2)$ for even $L$ is given under the condition $e^{2ik L}=1$, and hence it turns from a minimum between the two peaks at $k_l$ and $k_{l+1}$ for $\gamma <2\abs{\thop}$ into a maximum of the Fabry-Perot type for $\gamma>2\abs{\thop}$;
see the video 2 in the supplementary material.

\begin{figure}
\centering
\includegraphics[width=70mm]{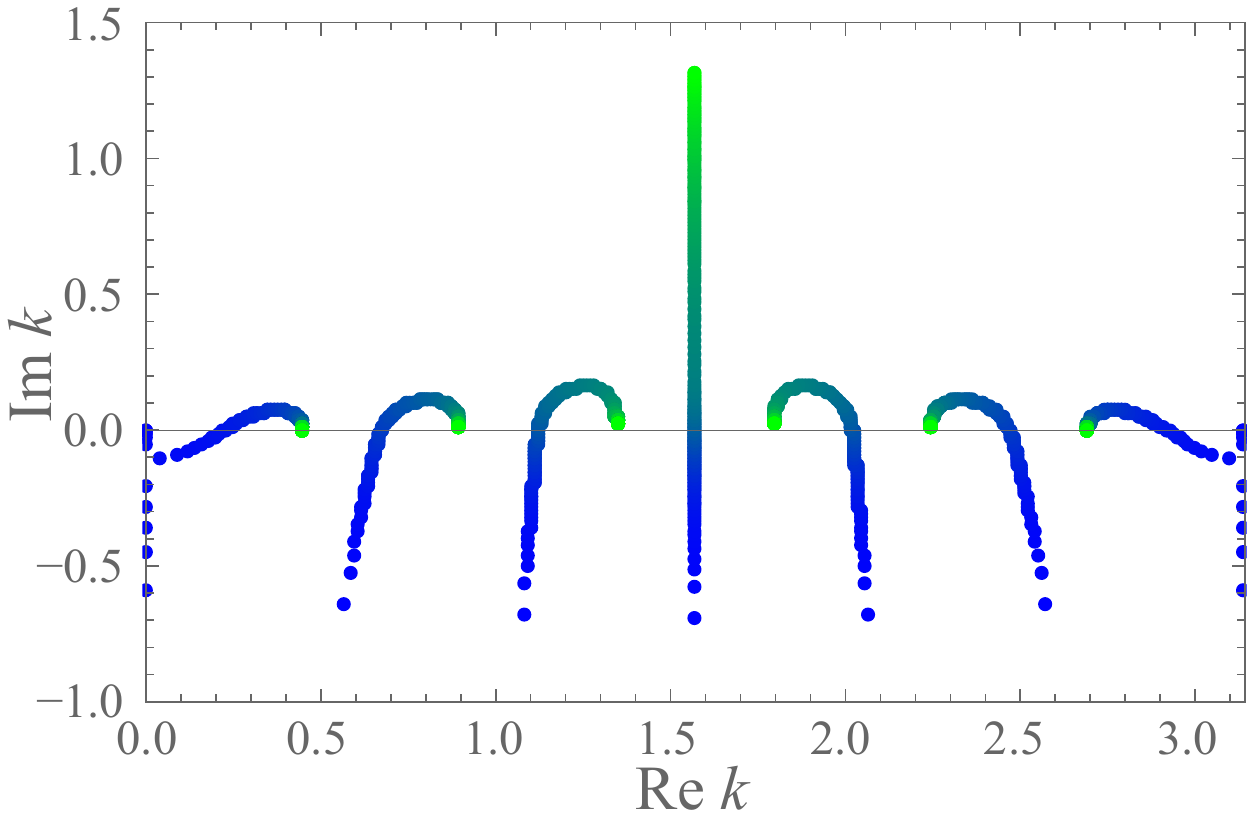}
\vspace{-\baselineskip}
\flushleft{(a)}\\
\vspace{\baselineskip}
\centering
\includegraphics[width=70mm]{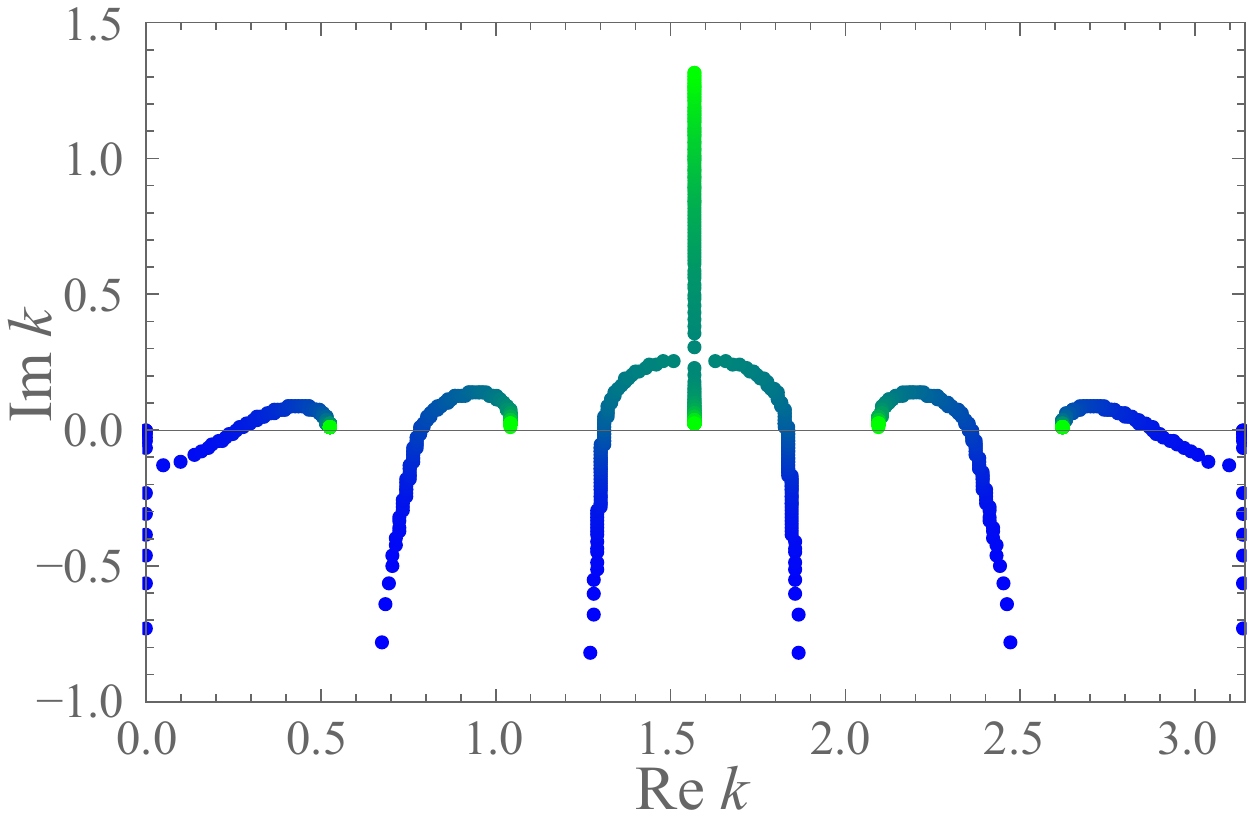}
\vspace{-\baselineskip}
\flushleft{(b)}\caption{Trajectories
of discrete eigenvalues in the complex $k$ plane.
Cases of (a)$L=7$, (b) $L=6$.
}
\label{k_sieg}
\end{figure}

\subsection{Discrete eigenvalues under the Siegert boundary condition}
\label{subsec-eig}

We can also understand the peak structure of the transmission probability from the locations of discrete eigenvalues.
The discrete eigenvalues are given by the poles of the $S$-matrix, and hence in the present case the zeros of the denominator of $\calT(k)$ and $\calR(k)$ in Eqs.~\eqref{FP-T} and~\eqref{FP-R}~\cite{landau}:
\begin{align}
4\thop^2\sin^2 k_n+\gamma^2(e^{2ik_nL}-1)=0.
\label{sieg_denom}
\end{align}
In the present case, there are $2L$ pieces of generally complex solutions $\{k_n\}$, except at exceptional points, where two solutions coallesce;
see App.~\ref{app:Siegert} for a review.
This implies that, except at the exceptional points, we may be able to break down $\calT(k)$ into the form of the Laurent expansion:
\begin{align}
\calT(k)=\sum_{n=1}^{2L-2} \frac{c_n}{k-k_n}+f(k),
\end{align}
where $\{c_n\}$ are generally complex constants and $f(k)$ is a regular function with the standard Taylor expansion.
Therefore, the most singular contributions to the transmission probability may be made of the Breit-Wigner-type ones~\cite{BW}:
\begin{align}
T(k)&=\abs{\calT(k)}^2
\simeq
\sum_{n=1}^{2L-2}
\frac{\abs{c_n}^2}{\abs{k-k_n}^2}
\nonumber\\
&=
\sum_{n=1}^{2L-2}
\frac{\abs{c_n}^2}{(k-\kr_n)^2+{\ki_n}^2};
\label{lorentzian}
\end{align}
that is, a Lorentzian peak with its center at $k=\kr_n$ and its half width at half maximum $\ki_n$.
Indeed, this is what we see in Figs.~\ref{T_L7} and~\ref{T_L6}.

Figure~\ref{k_sieg} shows the trajectories of the discrete eigenvalues in the complex $k$ plane on the side of $\Re k>0$ for $L=7$ and $L=6$;
see the video 3 in the supplementary material.
(There are $2L$ pieces of eigenvalues, but only half of them are on this side of the $k$ plane;
the remaining half are on the side of $\Re k<0$ and are irrelevant to the present argument.)

Upon increasing the parameter $\gamma$,
the eigenvalues leave trajectories that cross the real axis,
which is responsible for the diverging peaks in the regime of weak non-Hermiticity, $\gamma<2\abs{\thop}$.
When an eigenvalues is located on the real axis, we have $\ki=0$ in the Lorentzian~\eqref{lorentzian}, and hence the peak may diverge (except when the cancellation occurs with the numerator).
The two eigenvalues on the far right and the far left first cross the real axis, and thereby generate the first two diverging peaks.
The two neighboring eigenvalues next cross the real axis, generating the next two peaks.
The third pair of eigenvalues generate the third pair of peaks, and so on.
For odd values of $L$, in particular, the one in the middle generates the last divergence when it crosses the point $k=\pi/2$.
These all happen in the weak non-Hermiticity regime.

We stress here that in Hermitian systems the discrete eigenvalues are prohibited in the first quadrant~\cite{landau}.
In this sense, the divergence due to the eigenvalues crossing the real axis onto the first quadrant is a distinctively non-Hermitian  phenomenon.

In the strong non-Hermiticity regime, $\gamma>2\abs{\thop}$, the eigenvalues except the one in the middle approach the points specified in Eq.~\eqref{FPcond}, and generate the Fabry-Perot type peaks there.
In the case of $L=7$, for example, there are six such peaks, that is, one less than the diverging peaks in the weak non-Hermiticity regime, because the eigenvalue with $\kr=\pi/2$ keeps climbing up the complex $k$ plane and does not contribute to the Fabry-Perot peaks.
In the case of $L=6$, two eigenvalues closest to the point $\kr=\pi/2$ eventually collide in the middle, and one climbs up the $k$ plane but the other comes back to the real axis, again making peaks one less than the diverging peaks  in the weak non-Hermiticity regime.
The eigenvalues come to the real axis only in the limit of $\gamma\to\infty$, which is, however, a trivial limit with $T\equiv 0$ and $R\equiv 1$ with the exception of $T=1$ and $R=0$ at the points~\eqref{FPcond}.
The Fabry-Perot peaks do not diverge because Eq.~\eqref{sieg_denom} reduces to the Fabry-Perot condition $e^{2ik_n L}=1$ in the limit $\gamma\rightarrow\infty$, 
where the remaining first term in Eq.~\eqref{sieg_denom}
cancels the numerator of Eq.~\eqref{FP-T}, always giving a perfect transmission $\calT(k)=1$.

In the model~\eqref{H_tb} discussed in Sec.~\ref{sec:single}, the region of weak non-Hermiticity $\gamma<2\abs{\thop}$ falls on a superficially Hermitian region with two real eigenvalues, while that of strong non-Hermiticity $\gamma>2\abs{\thop}$  with two imaginary eigenvalues is considered to be a truly non-Hermitian region.
The roles of the two regimes seem to be consistent with the intuition that we have from the $2\times 2$ model~\eqref{H_PT2} in Introduction.
In our open \PT-symmetric system~\eqref{H_PT}, in contrast, real eigenvalues appear only occasionally in the regime $\gamma<2\abs{\thop}$, and the most eigenvalues approach the real axis again in the limit of extremely strong non-Hermiticity, which is a novel finding here.

The existence of the non-Hermitian Fabry-Perot region is specific to
the \PT-symmetric choice of $V_0$ and $V_L$, \textit{i.e.},
$V_0=i\gamma$ and $V_L=-i\gamma$.
In a non-Hermitian but non-\PT-symmetric choice of $V_0$ and $V_L$, \textit{e.g.},
$V_0=V_L=i\gamma$ or $V_0=V_L=-i\gamma$,
Fabry-Perot peaks such as the ones in the \PT-symmetric case
do not appear. In this sense
the non-Hermitian Fabry-Perot region is protected by the \PT-symmetry.
In the case of real (Hermitian) scattering potentials, \textit{e.g.},
$V_0=V_L=\gamma$ or $V_0=-V_L=\gamma$,
the Fabry-Perot peak structure appears for an arbitrary finite $\gamma$,
while
unlike in the \PT-symmetric case
the discrete eigenvalues approach the real $k$ axis from below in the limit of large $\gamma$.
In the \PT-symmetric case
the discrete eigenvalues approach the real $k$ axis from above;
see Fig.~\ref{k_sieg}.

\subsection{The reflection probability and the unitarity of the $S$-matrix}
\label{subsec:peakR}

Let us turn our attention to the reflection probability and analyze its peak structure based on the analytic formula~\eqref{FP-R}.
Since the denominator is the same as that of $\calT(k)$ in Eq.~\eqref{FP-T}, the divergent behavior is common in the regime $\gamma<2\abs{\thop}$.
The drastic difference is noticeable in the Fabry-Perot region $\gamma>2\abs{\thop}$.
The reflection probability $R(k)$ shows dips at $k$ given in Eq.~\eqref{FPcond},
corresponding to the Fabry-Perot peaks in $T(k)$; compare Figs.~\ref{T_L7}(b) and~\ref{R_L7}(a). 
In between the dips, however, $R(k)$ remains greater than the unity, which does not seem Hermitian even in the Fabry-Perot region $\gamma>2\abs{\thop}$.

We can understand this in the following way.
Equation~\eqref{FP-R} yields the reflection probability $R(k):=|\calR(k)|^2$ in the form
\begin{align}
R(k)=\frac{\gamma^2\qty(\gamma-2\thop\sin k)^2\sin^2 kL}%
{4\thop^4\sin^4 k + \gamma^2 (\gamma^2 - 4\thop^2\sin^2 k)\sin^2 kL}.
\label{Rk}
\end{align}
Comparing Eqs.~\eqref{Tk} and~\eqref{Rk},
we  find
\begin{align}
R(k)=\alpha(\gamma,k)\qty(1-T(k)),
\label{1-T}
\end{align}
where
\begin{align}
\alpha(\gamma,k):={\gamma-2\thop\sin k \over\gamma+2\thop\sin k}.
\label{alpha}
\end{align}
Therefore, $R(k)$ is related to $1-T(k)$, 
but is modulated as in Eq.~\eqref{1-T} 
by the function $\alpha(\gamma,k)$.
Since $\thop<0$ and $k>0$, we have $\alpha(\gamma,k)>1$ in the Fabry-Perot region $\gamma>2\abs{\thop}$,
which we can confirm in Fig.~\ref{R_L7}(a).

Figure~\ref{R_L7}(b), on the other hand, shows the reflection probability $\tilde{R}(k)$ due to the incident wave from the right.
This reflection probability shows dips at $k$ given in Eq.~\eqref{FPcond} too, but in between the dips, $\tilde{R}(k)$ is suppressed to values less than unity.

This is understand in the following way too.
As we stated at the end of Subsec.~\ref{subsec:FP}, we obtain the reflection probability $\tilde{R}(k)$ by flipping the sign of $\gamma$ from $R(k)$:
\begin{align}
\tilde{R}(k)=\frac{\gamma^2\qty(-\gamma-2\thop\sin k)^2\sin^2 kL}%
{4\thop^4\sin^4 k + \gamma^2 (\gamma^2 - 4\thop^2\sin^2 k)\sin^2 kL}.
\end{align}
Meanwhile, we have $\tilde{T}(k)=T(k)$. We therefore arrive at
\begin{align}
\tilde{R}(k)=\frac{1}{\alpha(\gamma,k)}\qty(1-\tilde{T}(k)).
\label{tildeR}
\end{align}
Therefore, 
$\tilde{R}(k)$ is related to $1-\tilde{T}(k)$, but is modulated by
the function $1/\alpha(\gamma,k)$, which 
is
less than unity in the Fabry-Perot region $\gamma>2\abs{\thop}$.
This is what we see in Fig.~\ref{R_L7}(b).

Let us finally analyze the sum of the transmission and reflection probabilities.
We obtain
\begin{align}
&T(k)+R(k)-1
\nonumber\\
&=\frac{-4\thop\gamma^2 \sin k \sin^2 kL(\gamma-2\thop \sin k)}%
{4\thop^4\sin^4 k + \gamma^2 (\gamma^2 - 4\thop^2\sin^2 k)\sin^2 kL}.
\label{T+R-1}
\end{align}
The denominator is non-negative because it is the square modulus of the common denominator of Eqs.~\eqref{FP-T} and~\eqref{FP-R}.
The numerator is always positive because 
we assume $\thop<0$, $\gamma>0$ and $k>0$.
We thereby conclude that $T(k)+R(k)>1$ 
always stands for the incident wave from the left.
For the incident wave from the right,
Eq.~\eqref{T+R-1} is changed to
\begin{align}
&\tilde{T}(k)+\tilde{R}(k)-1
\nonumber\\
&=\frac{-4\thop\gamma^2 \sin k \sin^2 kL(-\gamma-2\thop \sin k)}%
{4\thop^4\sin^4 k + \gamma^2 (\gamma^2 - 4\thop^2\sin^2 k)\sin^2 kL}.
\label{tilT+tilR-1}
\end{align}
We have $\tilde{T}(k)+\tilde{R}(k) <1$ in the Fabry-Perot regime because $\qty(-\gamma-2\thop\sin k)\leq\qty(-\gamma+2\abs{\thop})<0$ when we assume $\gamma>0$, $\thop<0$ and $k>0$.
Recall that the transmission  is symmetric with respect to the direction of the incident wave,
but the reflection  is not.

The difference between the two inequalities~\eqref{T+R-1} and~\eqref{tilT+tilR-1} is consistent with the intuition that the potential that the incident wave first meets makes larger contributions to the Fabry-Perot superposition~\eqref{FPser}:
since the incident wave from the left meets the source $+i\gamma$ first, the flux is enhanced; on the other hand, 
since the incident wave 
from the right meets the sink $-i\gamma$ first, the flux is suppressed.

\section{Scattering problem of the continuum model}
\label{sec:cont}

We finally describe what happens for the continuum model.
Let us  start with the tight-binding model, and consider its continuum limit.
We first recover the lattice constant $a$, and take the continuum limit
in which we set $a\rightarrow 0$ and $L\rightarrow\infty$,
keeping $\tilde{L}=La$ finite.
In this process, we replace the on-site scattering potential 
$\pm i\gamma$ with $\pm i\gamma a$ so that
they may converge to $\delta$-function scatterers $V(x) = i\gamma \delta (x)-i\gamma\delta(x-\tilde{L})$ in the continuum limit.
To reproduce the conventional kinetic term,
we also make the replacement: $-\thop a^2=\hbar^2/(2m)$.
Then, Eq.~\eqref{FP-T} reduces to
\begin{align}
\calT(k)=
{4k^2\over 4k^2+ \tilde{\gamma}^2(e^{2ik\tilde{L}}-1)},
\label{CALconti}
\end{align}
where we have introduced
\begin{align}
\tilde{\gamma}={\gamma\over \abs{\thop}}={2m\over \hbar^2}\gamma.
\label{tilde}
\end{align}
The same applies for Eq.~\eqref{FP-R}, producing
\begin{align}
\calR(k)={-\tilde{\gamma}(2k+\tilde{\gamma})(e^{2ik\tilde{L}}-1)
\over 4k^2 + \tilde{\gamma}^2(e^{2ik\tilde{L}}-1)}.
\label{BALconti}
\end{align}
The discrete eigenvalues are found from the zeros of the denominator:
\begin{align}
4k^2+\tilde{\gamma}^2\qty(e^{2ik\tilde{L}}-1)=0.
\end{align}
We can of course find the same expressions by solving the 
the Schr\"odinger equation $H\psi(x) = E\psi(x)$
with the Hamiltonian,
\begin{align}
H=-{\hbar^2\over 2m}{d^2\over dx^2}+i\gamma \delta(x) - i\gamma\delta(x-\tilde{L}).
\end{align}

\begin{figure}
\centering
\includegraphics[width=70mm]{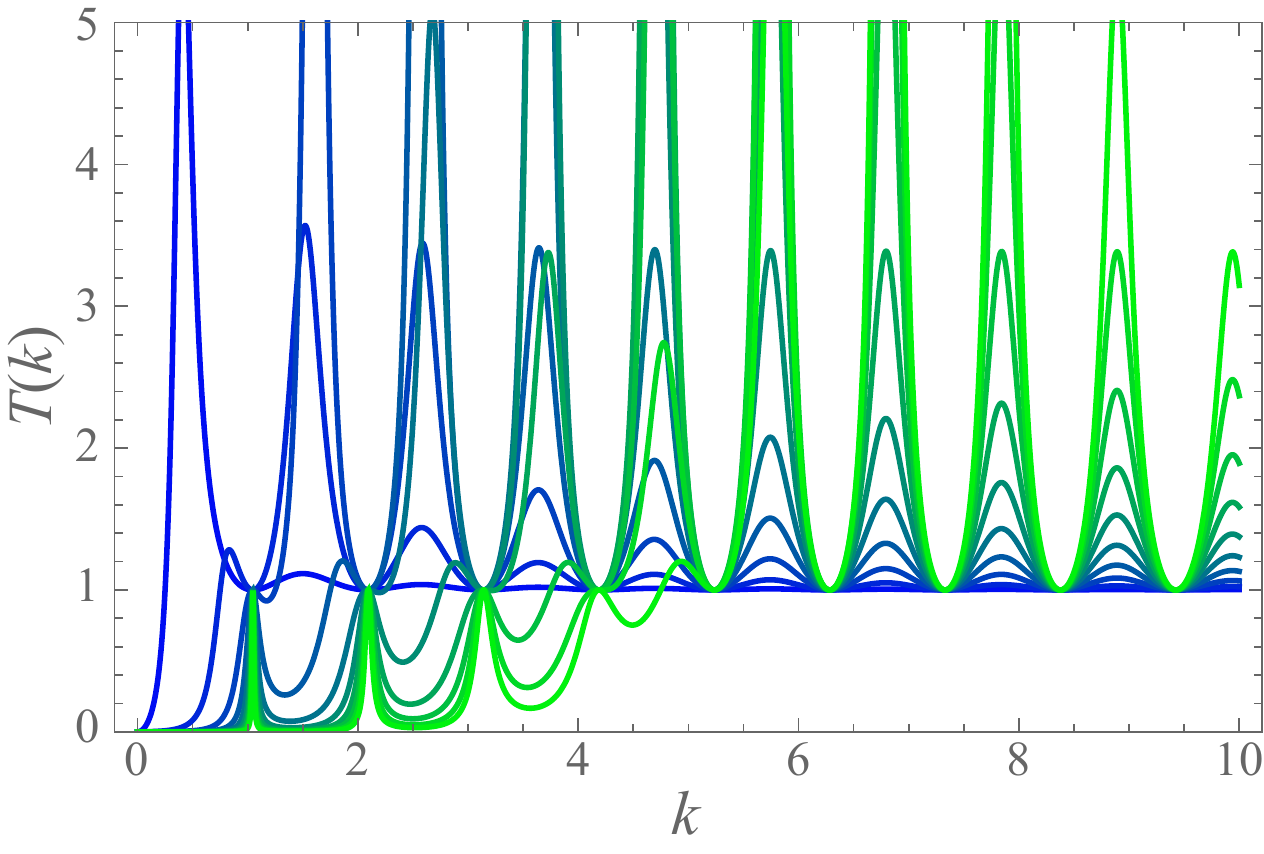}
\vspace{-\baselineskip}
\flushleft{(a)}\\
\vspace{\baselineskip}
\centering
\includegraphics[width=70mm]{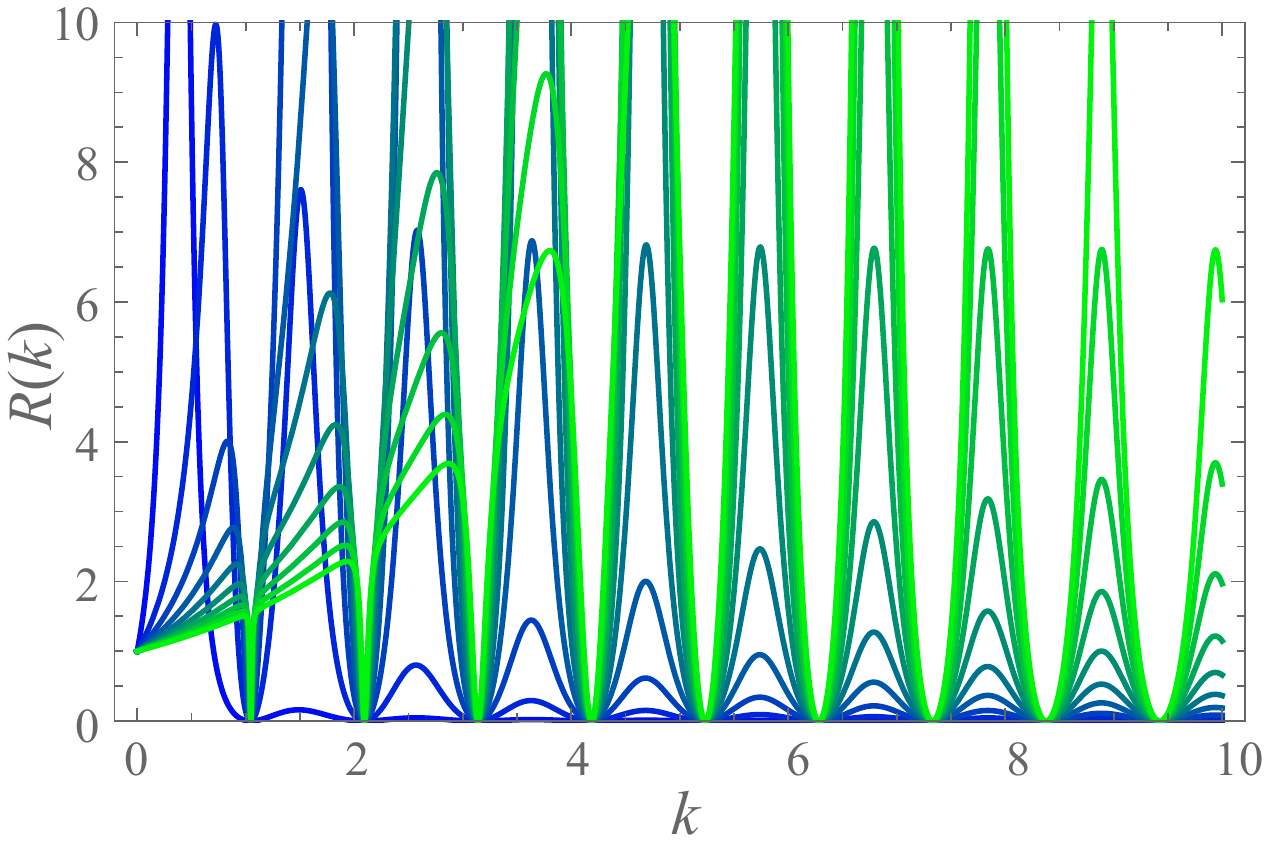}
\vspace{-\baselineskip}
\flushleft{(b)}\\
\vspace{\baselineskip}
\centering
\includegraphics[width=70mm]{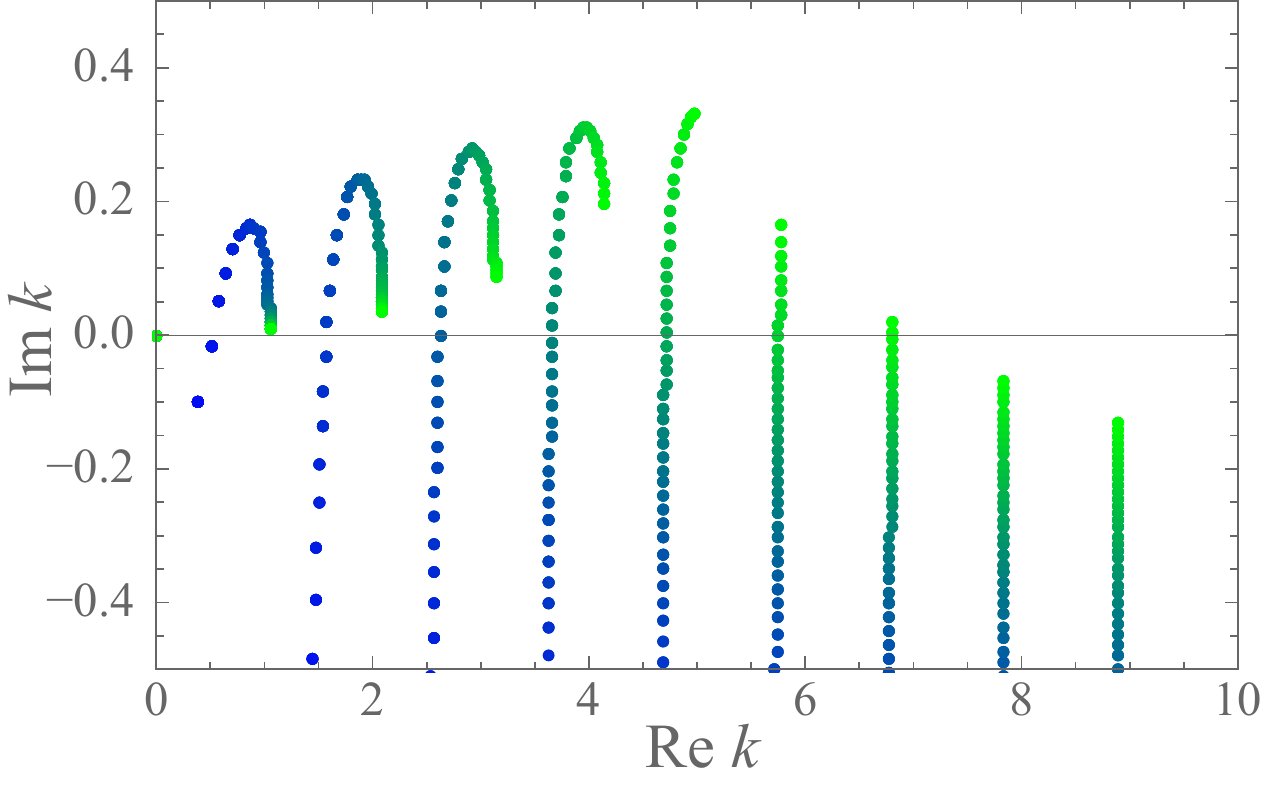}
\vspace{-\baselineskip}
\flushleft{(c)}\\
\caption{Variation of 
(a) the transmission probability $T(k)$ and
(b) the reflection probability $R(k)$ 
with $\gamma /\abs{\thop}$ varied from $0$ (blue) to $10$ (green) and $\tilde{L}=3$.
(c) The corresponding trajectories of discrete eigenvalues in the complex $k$ plane.
}
\label{conti}
\end{figure}

Figures~\ref{conti}(a) and (b) show, respectively,
the transmission and reflection probabilities in the continuum limit, in which
$\tilde{L}$ is chosen as $\tilde{L}=3$ so that
the separation of the Fabry-Perot peaks is $\Delta k=\pi/\tilde{L}\simeq 1$.
Figure~\ref{conti}(c) shows the trajectories of the discrete eigenvalues.
We can understand the features from the ones in Figs.~\ref{T_L7} and~\ref{T_L6} as follows.
We can regard the dispersion relation of the continuum model, namely $E(k)=\hbar^2k^2/(2m)=\abs{\thop} a^2 k^2$ as the leading term of the expansion of the lattice dispersion $E(k)=-2\abs{\thop}\cos ka$ with respect to the small parameter $a$ except for the constant shift $-2\abs{\thop}$.
In this sense, the low-energy region of the continuum model is magnification of the low-energy limit of the lattice model.
Therefore, what happens in Fig.~\ref{conti} magnifies what happens in the left half of each panel of Fig.~\ref{T_L7} with much more peaks.

We can thus see that for a fixed range of $k$, $T(k)$ and $R(k)$ diverge when the corresponding eigenvalue crosses the real axis for a relatively small value of $\tilde{\gamma}$. Then they turn to the Fabry-Perot-type behavior when the corresponding eigenvalue turns and comes back toward the real axis.
As a result, we observe the following:
\begin{enumerate}
\item 
a Fabry-Perot region appears in the range  $k<\tilde{\gamma}/2$;
\item 
divergent peaks appear in the intermediate range $k>\tilde{\gamma}/2$;
\item 
in the regime of $k\gg\tilde{\gamma}/2$,
\textit{i.e.}, at $k\rightarrow\infty$,
one recovers a unitary but trivial behavior. $T(k)=1$ and $R(k)=0$. 
\end{enumerate}
In the Fabry-Perot regime, $k<\tilde{\gamma}/2$,
the peaks appear when
\begin{align}
e^{2ik\tilde{L}}=1,
\end{align}
\textit{i.e.}, at $k=n\pi/\tilde{L}$ ($n=1,2,\cdots$), while
in the intermediate regime,
the divergence occurs at
\begin{align}
e^{2ik\tilde{L}}=-1,
\end{align}
\textit{i.e.}, at
\begin{align}
k=k^{(n)}={(2n-1)\pi\over 2\tilde{L}}\ \ (n=1,2,\cdots)
\label{kn}
\end{align}
and at the value of $\tilde{\gamma}$ such that
\begin{align}
\tilde{\gamma}=\sqrt{2}k^{(n)}.
\end{align}

\section{Concluding remarks}
\label{sec:conclusion}

We have considered the non-Hermitian scattering problem
for a Fabry-Perot-type \PT-symmetric model.
We found that both the transmission and reflection probabilities,
$T(k)$ and $R(k)$, behave strongly atypicaly
with divergent peaks
in the regime of weak non-Hermiticity $\gamma<2\abs{\thop}$,
while the behavior of $T(k)$
becomes superficially Hermitian
in the regime of strong non-Hermicitity $\gamma>2\abs{\thop}$.
In the latter, $T(k)$ shows conventional Fabry-Perot peaks that
are bounded by unity,
\textit{i.e.},
$T(k) \le 1$ as in the Hermitian case,
and yet
the behavior of $R(k)$ is unconventional,
leading to
breaking of the unitarity of the $S$-matrix:
$T^2+R^2 \neq 1$.

We exactly obtained the expressions for the transmission and reflection coefficients $\calT(k)$ and $\calR(k)$
by simply summing up the Fabry-Perot infinite series.
By interpreting this formula, we have clarified
the reason why
$T(k)$ and $R(k)$ drastically changes their behavior at $\gamma=2\abs{\thop}$.

\begin{acknowledgements}
The authors thank Hideaki Obuse for helpful discussions at an early stage of the present work.
K.S., K.K. and K.I. thank Y. Kadoya, M. Nishida, A. Tanaka and A. Kimura for useful comments and discussions.
N.H.'s work was supported by JSPS KAKENHI Grant Number 19H00658.
K.I. has been supported by JSPS KAKENHI Grant Number 20K03788 and 18H03683.
\end{acknowledgements}

\appendix

\section{Discrete eigenvalues of open quantum systems under the Siegert boundary condition}
\label{app:Siegert}

Once we analytically continue the transmission and reflection coefficients~\eqref{FP-T} and~\eqref{FP-R} onto the complex $k$ plane, we find poles, which are often termed resonance poles.
In fact, one of the textbook definitions of resonance is a pole of the $S$-matrix, but we can also define it as an eigenstate of the Schr\"{o}dinger equation under a specific boundary condition.

Since the transmission and reflection coefficients have the denominator $A$, their poles are given by the zeros of $A$~\cite{landau}.
This means that the wave function at the resonance poles are given by putting $A$ to zero in Eq.~\eqref{inc-trm}, that is,
\begin{align}
\psi_x=\begin{cases}
Be^{-ikx} & \mbox{for $x\le 0$},\\
Ce^{ikx} & \mbox{for $x\ge L$}.
\end{cases}
\label{SiegertBC}
\end{align}
This wave function contains out-going waves only for $\Re k>0$ and in-coming waves only for $\Re k<0$, which is called the Siegert boundary condition~\cite{sieg1} for the wave functions of discrete eigenvalues; 
see Ref.~\onlinecite{hatano08} for a summary.

In fact, the solutions include all kinds of discrete states with point spectra, namely, bound states, anti-bound states, resonant states and anti-resonant states.
To give an example, the wave function of the form~\eqref{SiegertBC} is a bound state if $k$ is a pure imaginary number with a positive imaginary part.
In the standard Hermitian scattering problem with the time-reversal symmetry, the solutions on the positive imaginary axis are the bound states as exemplified above, those on the negative imaginary axis are called the anti-bound states, those in the fourth quadrant of the complex $k$ plane are the resonant states, and those in the third quadrant are called the anti-resonant states, which are the time-reversal states of the resonant states. (There are continuum states on the real axis, in addition.) 
States on the upper half of the complex $k$ plane are prohibited except on the positive imaginary axis,  because of the normalization~\cite{landau}.
For non-Hermitian problems, however, it is known that the poles can move across the real axis of the complex $k$ plane~\cite{garmon}.

In the specific example of the model~\eqref{H_PT}, we obtain the eigenvalue equation for the Siegert boundary condition~\eqref{SiegertBC} by setting $\tilde{A}$ to zero in Eq.~\eqref{ML1}, and therefore the discrete eigenvalues are the solutions of $\det M_L=0$, which we find from Eq.~\eqref{detM2}:
\begin{align}
4\thop^2\sin^2 k-\gamma^2\qty(e^{2ikL}-1)=0.
\label{res-eq1}
\end{align}
This is nothing but the equation for the zeros of the denominator of $\calT(k)$ and $\calR(k)$ in Eqs.~\eqref{FP-T} and~\eqref{FP-R}.

For numerical calculations of the discrete eigenvalues, however, finding all solutions of the nonlinear equation~\eqref{res-eq1} is generally not easy.
We here briefly review a convenient method of numerically finding the discrete eigenvalues; see Ref.~\onlinecite{hatano14} for details.
The eigenvalue equation to be solved reads
\begin{align}
\mqty(
i\gamma+\thop e^{ik}  &  \thop &&& \\
\thop &  & \thop && \\
& \thop &  &  \ddots & \\
 && \ddots &  &  \thop \\
&&& \thop  &  -i\gamma+\thop e^{ik} 
)
\vec{\psi}(k)
=E\vec{\psi}(k).
\label{nonlinear-eig}
\end{align}
This is a nonlinear eigenvalue problem because the left-hand side is a function of the eigenvalue $E$ through the wave number $k=\arccos(E/\thop)$.

More specifically, we can cast it into a second-order eigenvalue problem with respect to $\beta=e^{ik}$.
Using this variable, we transform Eq.~\eqref{nonlinear-eig} to
\begin{align}
\qty(\beta^2 U+\beta V +W)\vec{\psi}(\beta)=0,
\label{nonlinear-eig1}
\end{align}
where
\begin{align}
U&=-\thop I_{L+1} 
+\thop \mqty(1 & & & & \\
& 0 & & & \\
& & \ddots & & \\
& & & 0 & \\
& & & & 1
),
\\
V&=\mqty(
i\gamma  &  \thop &&& \\
\thop &  & \thop && \\
& \thop &  &  \ddots & \\
&& \ddots &  &  \thop \\
&&& \thop  &  -i\gamma
),
\\
W&=-\thop I_{L+1}.
\end{align}
We can further transform this into a generalized but linear eigenvalue equation by doubling the vector space as follows~\cite{secondorder,hatano14}:
\begin{align}
\mqty(
\beta I_{L+1} & -I_{L+1} \\
W & \beta U+V 
)
\mqty(\vec{\psi}(\beta)\\
\beta\vec{\psi}(\beta)
)=0,
\label{double}
\end{align}
which is a $(2L+2)$-dimensional matrix equation.
The $(L+1)$-dimensional first row of Eq.~\eqref{double} guarantees that the second row of the vector is always $\beta$-fold its first row.
The second row of the equation is equivalent to Eq.~\eqref{nonlinear-eig1}.

Equation~\eqref{double} is a linear eigenvalue equation with respect to $\beta$ in the sense that
\begin{align}
\mqty(0 & I_{L+1} \\
-W & -V )\mqty(\vec{\psi}(\beta)\\
\beta\vec{\psi}(\beta)
)
=\beta \mqty( I_{L+1} & 0 \\
0 & U )\mqty(\vec{\psi}(\beta)\\
\beta\vec{\psi}(\beta)
).
\end{align}
We can numerically find the $2(L+1)$ pieces of eigenvalues for small $L$ easily.
Note, however, that because the $(1,1)$- and $(L+1,L+1)$-elements of $U$ are missing, we in fact find only $2L$ pieces of eigenvalues in the present case.
In order to find full eigenvalues, we should introduce modulation of the hopping amplitudes on the left and right edges of the scattering region $[0,L]$; see Appendix H of Ref.~\onlinecite{sasada11}.

\section{Analytic expressions of the transmission and reflection probabilities}
\label{app:ML}

We here invert the matrix $M_L$ in Eq.~\eqref{ML2} and obtain the formulas~\eqref{FP-T} and~\eqref{FP-R} from the expressions~\eqref{MI-B/A} and~\eqref{MI-C/A}.
The $(i,j)$-element of the inverse matrix ${M_L}^{-1}$ is given by
\begin{align}
\qty({M_L}^{-1})_{ij}=(-1)^{i+j}\frac{{\det\,}{M_L}^{ji}}{\det M_L} .
\label{MLinv}
\end{align}
where
${\det\,}{M_L}^{ji}$ is the cofactor of $M_L$, that is, the determinant of an $L\times L$  matrix ${M_L}^{ji}$ that we make from the $(L+1)\times (L+1)$ matrix $M_L$ by removing the $j$th row and the $i$th column.
In order to write down the expressions~\eqref{MI-B/A} and~\eqref{MI-C/A} explicitly, we therefore need $\det M_L$ as well as the cofactors ${\det\,}{M_L}^{1,L+1}$ and ${\det\,}{M_L}^{1,1}$.

We can find the determinant of the matrix $M_L$ in Eq.~\eqref{ML2} by cofactor expansion.
For brevity of the notation, let us fix $L=4$ for the moment.
The cofactor expansion with respect to the first row gives
\begin{align}
\det M_4&=
\qty(i\gamma - \thop e^{-ik})\times
\nonumber\\
&\quad\det\mqty(
-E(k) & \thop & & \\
\thop & -E(k) & \thop & \\
&\thop & -E(k) & \thop \\
&& \thop & -i\gamma -\thop e^{-ik}
)
\nonumber\\
&-\thop^2
\det\mqty(
 -E(k) & \thop & \\
\thop & -E(k) & \thop \\
 & \thop & -i\gamma -\thop e^{-ik}
).
\end{align}
The cofactor expansion further with respect to the last row gives
\begin{align}
\det M_4 &=\qty(i\gamma - \thop e^{-ik})\qty(-i\gamma - \thop e^{-ik})\times
\nonumber\\
&\quad\det\mqty(
-E(k) & \thop & \\
\thop & -E(k) & \thop  \\
&\thop & -E(k) &  \\
)
\nonumber\\
&-\qty(i\gamma - \thop e^{-ik})\thop^2
\det\mqty(
-E(k) & \thop \\
\thop & -E(k)
)
\nonumber\\
&-\thop^2\qty(-i\gamma - \thop e^{-ik})
\det\mqty(
-E(k) & \thop \\
\thop & -E(k)
)
\nonumber\\
&+\thop^4\det\mqty(-E(k)).
\end{align}
This algebra for $L=4$ implies the general expression
\begin{align}
&\det M_L
\nonumber\\
&=(\gamma^2+\thop^2e^{-2ik})d_{L-1}+2\thop^3e^{-ik}d_{L-2}+\thop^4d_{L-3},
\label{ML3}
\end{align}
where 
$d_n$ denotes the determinant of $n\times n$ matrix with all the diagonal elements $-E(k)=-\thop(e^{ik}+e^{-ik})$ and all super- and sub-diagonal elements $\thop$.

We can find $d_n$ again by cofactor expansion to obtain the recurrence equation
\begin{align}
d_n
&=-\thop(e^{ik}+e^{-ik}) d_{n-1}-\thop^2 d_{n-2}.
\label{dn}
\end{align}
The solutions of the characteristic equation for the recurrence equation~\eqref{dn}, $\xi^2+2\xi \thop \cos k+\thop^2=0$, are $\xi=-\thop e^{\pm ik}$, and hence we have
\begin{align}
d_n+\thop \beta d_{n-1}&=-\thop \beta^{-1}\qty(d_{n-1}+\thop \beta d_{n-2})
\\
&=\qty(-\thop)^{n-2}\beta^{-n+2}\qty(d_2+\thop \beta d_1)
\nonumber\\
&=\qty(-\thop)^n \beta^{-n},
\end{align}
where $\beta=e^{ik}$.
Transforming this to
\begin{align}
\frac{d_n\beta^n}{\qty(-\thop)^n}=\beta^2\frac{d_{n-1}\beta^{n-1}}{\qty(-\thop)^{n-1}}+1
\end{align}
gives
\begin{align}
\frac{d_n\beta^n}{\qty(-\thop)^n}+\frac{1}{\beta^2-1}
&=\beta^2\qty[\frac{d_{n-1}\beta^{n-1}}{\qty(-\thop)^{n-1}}+\frac{1}{\beta^2-1}]
\\
&=\beta^{2(n-1)}\qty(\frac{d_1\beta}{-\thop}+\frac{1}{\beta^2-1})
\nonumber\\
&=\frac{\beta^{2(n+1)}}{\beta^2-1},
\end{align}
and hence
\begin{align}
d_n=\qty(-\thop)^n\frac{\beta^{n+2}-\beta^{-n}}{\beta^2-1}
=\qty(-\thop)^n\frac{\sin(n+1)k}{\sin k}.
\end{align}
Substituting this into Eq.~\eqref{ML3}, we finally arrive at
\begin{align}
\det M_L
&=\qty(-\thop)^{L+1}
\qty[\tilde{\gamma}^2\frac{\beta^{L}-\beta^{-L}}{\beta-\beta^{-1}}
-\beta^{-L}(\beta-\beta^{-1})]
\nonumber\\
&=\qty(-\thop)^{L+1}\qty(\tilde{\gamma}^2\frac{\sin kL}{\sin k}
-2ie^{-ikL}\sin k),
\label{detM2}
\end{align}
where $\tilde{\gamma}=g/\thop$.

Let us next compute the cofactors ${\det\,}{M_L}^{1,L+1}$ and ${\det\,}{M_L}^{1,1}$.
The easier is the former:
\begin{align}
&{\det\,}  {M_L}^{1,L+1}
=\thop^L.
\label{num1}
\end{align}
The slightly more complicated is the latter. 
By cofactor expansion, we have
\begin{align}
{\det\,} {M_L}^{1,1}&=
(-i\gamma-\thop\beta^{-1})d_{L-1}-\thop^2 d_{L-2}
\nonumber\\
&=\qty(-\thop)^L
\qty(i\tilde{\gamma}\frac{\beta^L-\beta^{-L}}{\beta-\beta^{-1}}
+\beta^{-L})
\nonumber\\
&=\qty(-\thop)^L
\qty(i\tilde{\gamma}\frac{\sin kL}{\sin k}
+e^{-ikL}).
\label{num2}
\end{align}

Summarizing the results~\eqref{detM2}--\eqref{num2}, we arrive at the two elements of the inverted matrix that we need as follows:
\begin{align}
\qty({M_L}^{-1})_{L+1,1}
&=
(-1)^{L+2}\frac{{\det\,}{M_L^{1,L+1}}}{\det M_L}
\nonumber\\
&=\frac{-2ie^{ikL}\thop\sin k}%
{4\thop^2 \sin^2 k+ \gamma^2 \qty(e^{2ikL}-1)},
\\
\qty({M_L}^{-1})_{1,1}
&=\frac{{\det\,}{M_L^{1,1}}}{\det M_L}
\nonumber\\
&=\frac{-2i\thop\sin k-i\gamma\qty(e^{2ikL}-1)}%
{4\thop^2 \sin^2 k+ \gamma^2 \qty(e^{2ikL}-1)}.
\end{align}
Inserting these into Eqs.~\eqref{MI-B/A} and~\eqref{MI-C/A}, we arrive at the same expressions as Eqs.~\eqref{FP-T} and~\eqref{FP-R}.

\bibliography{FPR_ref}

\end{document}